\pdfoutput=1
\documentclass[11pt,twoside,a4paper,cmspaper,final,collab]{cms-tdr}
\def\svnVersion{bdaa400}\def\svnDate{2024/06/05}\def\cmsCernNoTag{CERN-EP-2023-144}\def\cmsCernDate{\today}\def\cmsMessage{Published in the Journal of High Energy Physics as \href{http://dx.doi.org/10.1007/JHEP05(2024)311}{\doi{10.1007/JHEP05(2024)311}.}}
\begin{document}\cmsNoteHeader{EXO-19-016}

\newcommand{\tautau}  {\ensuremath{\PGt\PGt}\xspace}
\newcommand{\mumu}    {\ensuremath{\PGm\PGm}\xspace}
\newcommand{\emu}     {\ensuremath{\Pe\PGm}\xspace}
\newcommand{\etau}    {\ensuremath{\Pe\tauh}\xspace}
\newcommand{\mutau}   {\ensuremath{\PGm\tauh}\xspace}
\newcommand{\ltau}    {\ensuremath{\Pell\tauh}\xspace}
\newcommand{\ditau}   {\ensuremath{\tauh\tauh}\xspace}
\newcommand{\LQ}      {\ensuremath{\mathrm{LQ}}\xspace}
\newcommand{\LQs}     {\ensuremath{\mathrm{LQs}}\xspace}
\newcommand{\mLQ}     {\ensuremath{m_{\LQ}}\xspace}
\newcommand{\mvis}    {\ensuremath{m_\text{vis}}\xspace}
\newcommand{\STMET}   {\ensuremath{S_\mathrm{T}^\text{MET}}\xspace}
\newcommand{\DYjets}  {\ensuremath{\text{DY}+\text{jets}}\xspace}
\newcommand{\Wjets}   {\ensuremath{\PW+\text{jets}}\xspace}
\newcommand{\Vgjets}  {\ensuremath{\PV\PGg+\text{jets}}\xspace}
\newcommand{\jtf}     {\ensuremath{j\to\tauh}\xspace}
\newcommand{\sigmafit}{\ensuremath{\sigma_\text{fit}}\xspace}
\newcommand{\muR}     {\ensuremath{\mu_\mathrm{R}}\xspace}
\newcommand{\muF}     {\ensuremath{\mu_\mathrm{F}}\xspace}
\newcommand{\geqb}    {\ensuremath{\geq}1b\xspace}
\newcommand{\pp}      {\ensuremath{\Pp\Pp}\xspace}
\newcommand{\Zee}     {\ensuremath{\PZ \to \Pe\Pe}\xspace}
\newcommand{\Zeepm}   {\ensuremath{\PZ \to \Pep\Pem}\xspace}
\newcommand{\Zmmpm}   {\ensuremath{\PZ \to \PGmp\PGmm}\xspace}
\newcommand{\DeepTau} {\textsc{DeepTau}\xspace}
\newlength\cmsTabSkip\setlength{\cmsTabSkip}{1ex}
\newlength\cmsFigWidth\setlength\cmsFigWidth{0.44\textwidth}

\cmsNoteHeader{EXO-19-016}
\title{Search for a third-generation leptoquark coupled to a \texorpdfstring{\PGt}{tau} lepton and a \texorpdfstring{\PQb}{b} quark through single, pair, and nonresonant production in proton-proton collisions at \texorpdfstring{$\sqrt{s}=13\TeV$}{sqrt(s) = 13 TeV}}

\date{\today}

\abstract{A search is presented for a third-generation leptoquark (\LQ) coupled exclusively to a \PGt lepton and a \PQb quark. The search is based on proton-proton collision data at a center-of-mass energy of 13\TeV recorded with the CMS detector, corresponding to an integrated luminosity of 138\fbinv. Events with \PGt leptons and a varying number of jets originating from \PQb quarks are considered, targeting the single and pair production of \LQs, as well as nonresonant $t$-channel \LQ exchange. An excess is observed in the data with respect to the background expectation in the combined analysis of all search regions. For a benchmark \LQ mass of 2\TeV and an \LQ-\PQb-\PGt coupling strength of 2.5, the excess reaches a local significance of up to 2.8 standard deviations. Upper limits at the 95\% confidence level are placed on the \LQ production cross section in the \LQ mass range 0.5--2.3\TeV, and up to 3\TeV for $t$-channel \LQ exchange. Leptoquarks are excluded below masses of 1.22--1.88\TeV for different \LQ models and varying coupling strengths up to 2.5. The study of nonresonant \tautau production through $t$-channel \LQ exchange allows lower limits on the \LQ mass of up to 2.3\TeV to be obtained.}

\hypersetup{
pdfauthor={CMS Collaboration},
pdftitle={Search for a third-generation leptoquark coupled to a tau lepton and a b quark through single, pair, and nonresonant production in proton-proton collisions at sqrt(s) = 13 TeV},
pdfsubject={CMS},
pdfkeywords={CMS, leptoquark, tau, bottom, third-generation, Beyond Standard Model}
}

\maketitle

\section{Introduction} \label{sec:introduction}

The standard model (SM) of particle physics is remarkably successful in describing matter and its fundamental interactions.
In 2012, the discovery of a major missing piece, the Higgs boson, was announced by the ATLAS and CMS experiments~\cite{Higgs_discovery_2012_ATLAS, Higgs_discovery_2012_CMS, Higgs_discovery_2013_CMS} at the CERN LHC.
Despite the many successes of the SM, there are both theoretical and experimental reasons why it is not considered to be the final and most fundamental theory of nature.
Measurements of the $\PB \to \PD \PGt \PGn$ and $\PB \to \PDst \PGt \PGn$ decay rates reported by the BaBar~\cite{RD_BaBar1,RD_BaBar2}, Belle~\cite{RD_Belle1,RD_Belle2,RD_Belle3,RD_Belle4,RD_Belle5,RD_Belle6}, and LHCb~\cite{RD_LHCb1,RD_LHCb2,RD_LHCb3,LHCb:2023zxo,LHCb:2023uiv} Collaborations collectively deviate from the SM predictions by about three standard deviations~\cite{HeavyFlavorAveragingGroup:2022wzx}.
Leptoquark (LQ) models have received considerable interest in the interpretation of these deviations.
Leptoquarks are hypothetical color-triplet scalar or vector bosons, which carry both baryon and lepton quantum numbers, and have fractional electric charge.
They are predicted by many theories beyond the SM, such as grand unification~\cite{GUT1,GUT2,GUT3,GUT4,GUT5,GUT6,GUT7,GUT8} and models invoking technicolor~\cite{technicolor1,technicolor2,technicolor3} or compositeness~\cite{compositeness1,compositeness2,compositeness3}.
An \LQ that couples most strongly to third-generation fermions could offer coherent explanations~\cite{RD_LQ_pheno1,RD_LQ_pheno2,RD_LQ_pheno3,RD_LQ_pheno4,RD_LQ_pheno5,RD_LQ_pheno6,RD_LQ_pheno7,RD_LQ_pheno8,RD_LQ_pheno9,RD_LQ_pheno10,RD_LQ_pheno11,RD_LQ_pheno12,RD_LQ_pheno13,RD_LQ_pheno14,RD_LQ_pheno15,RD_LQ_pheno16,RD_LQ_pheno17,RD_LQ_pheno18,RD_LQ_pheno19,RD_LQ_pheno20,RD_LQ_pheno21,RD_LQ_pheno22,RD_LQ_pheno23,RD_LQ_pheno24,RD_LQ_pheno25,RD_LQ_pheno26,RD_LQ_pheno27,RD_LQ_pheno28,RD_LQ_pheno29,RD_LQ_pheno30,RD_LQ_pheno31,RD_LQ_pheno32} for the $\PB \to \PD \PGt \PGn$, $\PB \to \PDst \PGt \PGn$, and other \PB physics anomalies reported by LHCb~\cite{B-Kmumu_LHCb1,B-Kmumu_LHCb2,B-phimumu_LHCb1,P5_LHCb1,P5_LHCb2,P5_LHCb3} and Belle~\cite{P5_Belle1}.

The production cross sections in proton-proton (\pp) collisions and decay widths of \LQs are determined by: the \LQ mass \mLQ, its branching fraction $\beta$ to a charged lepton and a quark, the coupling strength $\lambda$ of the \LQ-lepton-quark vertex, and the coupling parameter $\kappa$ in the case of vector \LQs~\cite{LQ_theory3,Blumlein:1996qp,LQ_theory_NLO}.
The \LQ can be produced in pairs via gluon fusion or quark-antiquark annihilation, or singly via quark-gluon fusion.
However, for large values of the coupling strength $\lambda$, the nonresonant production of two leptons via \LQ exchange in the $t$ channel dominates the total cross section~\cite{LQ_tchannel,LQ_large_coupling}.
Pair production of \LQs is approximately independent of $\lambda$, while single production and nonresonant production are not.

Currently, the most stringent limits on the production of an \LQ decaying to a \PGt lepton and a \PQb quark come from analyses by the ATLAS Collaboration~\cite{ATLAS:2023uox, ATLAS:2023vxj}. Leptoquarks with masses below 1.49--1.96\TeV were excluded in a search for scalar or vector \LQ pair production~\cite{ATLAS:2023uox}, while coupling-dependent limits have been placed in a search for single and nonresonant \LQ production~\cite{ATLAS:2023vxj}, excluding scalar (vector) \LQ masses of up to 1.53 (2.05)\TeV at couplings of $\lambda = 2.5$.
A search for single production of \LQs with the same decay mode was performed by the CMS Collaboration, excluding scalar \LQs below masses of 1.05\TeV for $\lambda = 2.5$~\cite{LQ3_CMS_btautau}.
Other searches for \LQs decaying to a \PQt quark and a \PGt lepton or neutrino exploiting the LHC 2016--2018 (2015--2018) data set have been performed by CMS~\cite{LQ3_tnubtau_CMS_Run2_SUSY,LQ3_tnubtau_CMS_Run2} (ATLAS~\cite{LQ3_bnuttau_ATLAS_Run2,LQ3_tnubtau_ATLAS_Run2}).

\begin{figure}[ht!]
  \centering
{\includegraphics[width=0.26\textwidth]{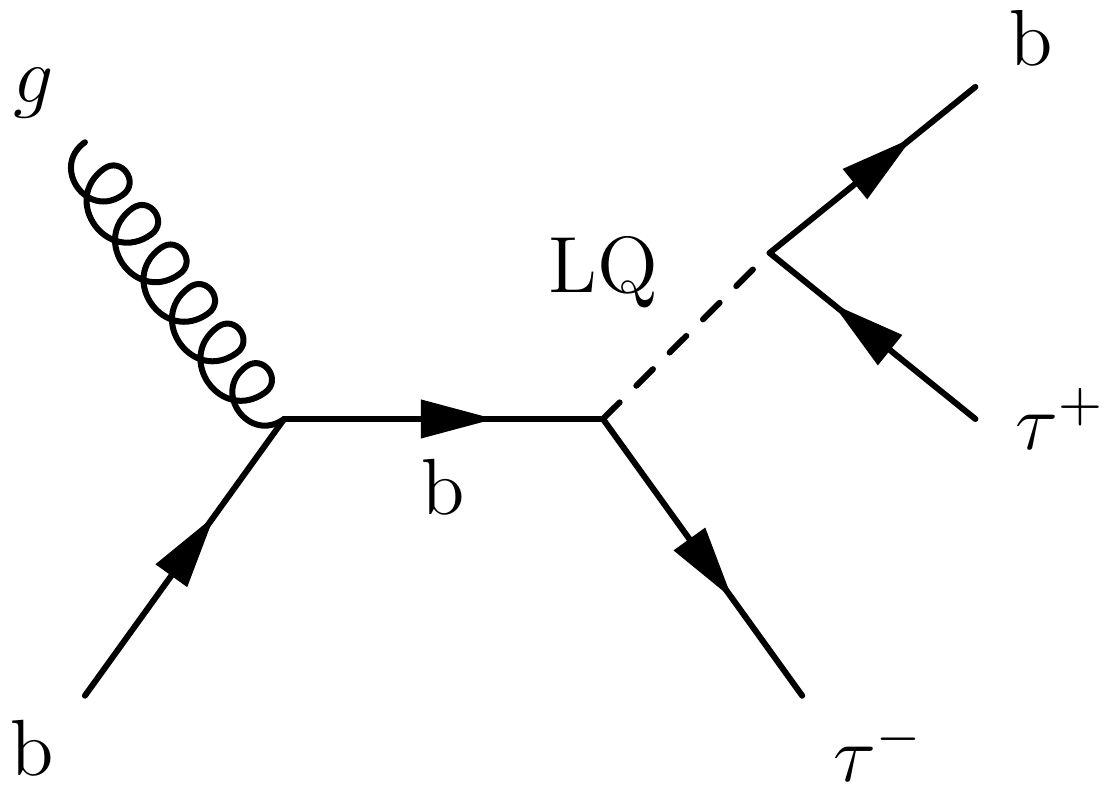}}
  \quad
{\includegraphics[width=0.25\textwidth]{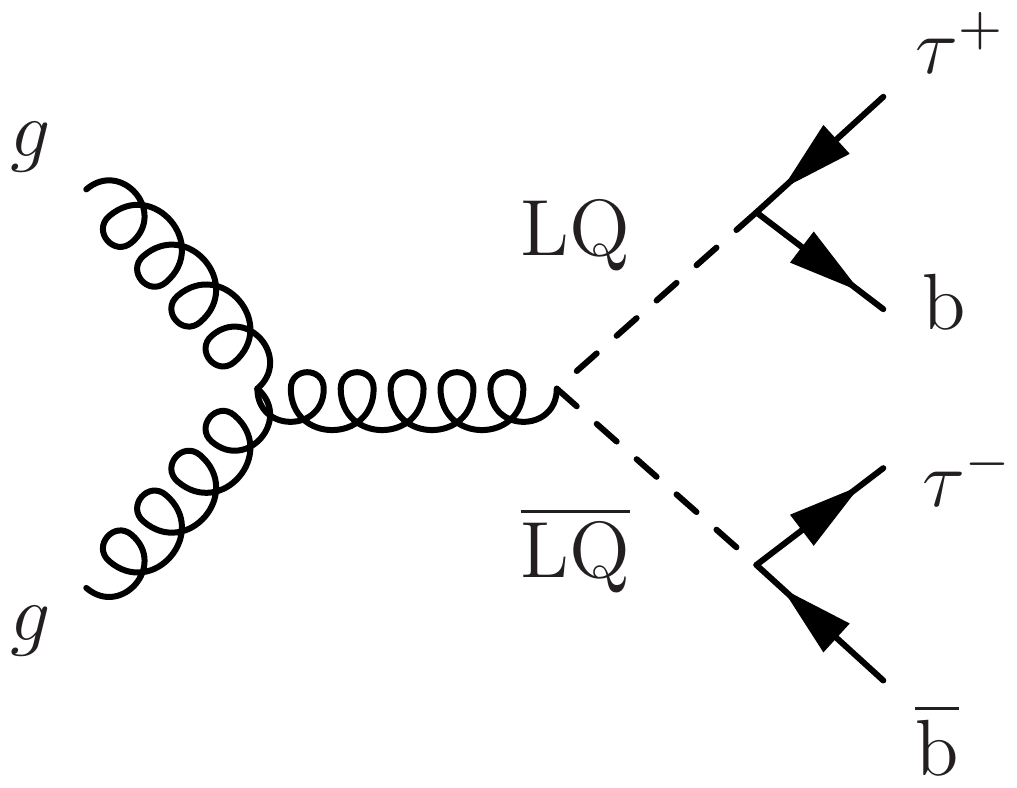}}
  \quad
{\includegraphics[width=0.21\textwidth]{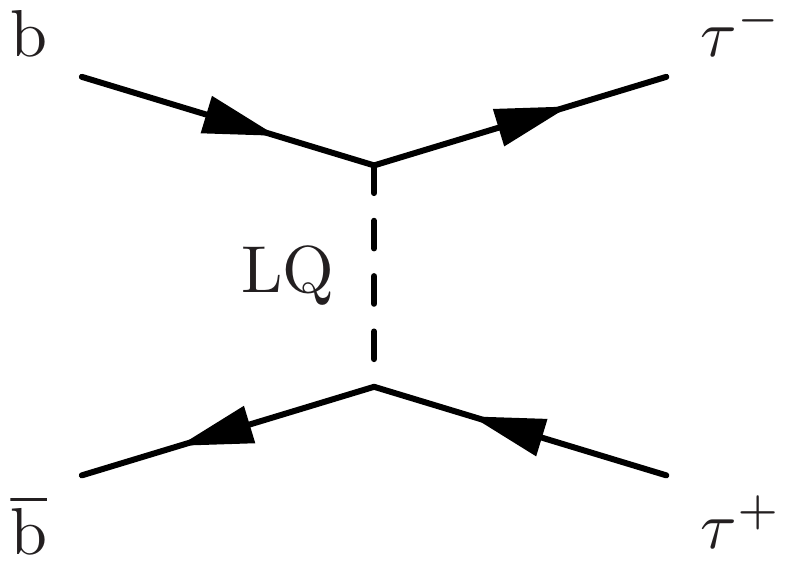}}
  \caption{
Example Feynman diagrams of signal processes at leading order: single (left) and pair (center) \LQ production,
as well as nonresonant production of two \PGt leptons via $t$-channel \LQ exchange (right).
  }
  \label{fig:diagrams}
\end{figure}

This paper presents a search that targets the single and pair production of scalar and vector \LQs that decay exclusively to a \PGt lepton and a \PQb quark, and a search for the nonresonant production of a \PGt lepton pair. Example Feynman diagrams of the signal processes at leading order (LO) are shown in Fig.~\ref{fig:diagrams}.
Single and pair production have a resonant final state that includes two \PGt leptons and at least one heavy-flavored jet. Nonresonant production will also have a final state with two \PGt leptons, and potentially additional jets originating from gluon splitting or initial-state radiation.

The search is based on a data sample of \pp collisions at a center-of-mass energy of 13\TeV recorded by the CMS experiment in 2016--2018, corresponding to an integrated luminosity of 138\fbinv. It exploits several decay channels of the \PGt lepton pair: fully hadronic (\ditau), semileptonic (\etau, \mutau), and fully leptonic channels (\emu, \mumu). Throughout this paper, electrons and muons are collectively referred to as leptons (\Pell). Owing to the large branching fraction of the \tauh decay, the semileptonic and fully hadronic channels are the most sensitive to an \LQ signal, while events with two leptons are primarily used to constrain the SM backgrounds.
Candidate events are further categorized according to whether or not at least one high-\PT jet is present (targeting resonant \LQ production) or absent (targeting nonresonant \tautau production). If high-\PT jets are present, two subcategories are defined according to whether or not at least one is identified as being initiated by a \PQb quark. The \LQ signal is extracted from a simultaneous fit to the observed distribution of the scalar sum of transverse momenta in the categories with a high-\PT jet, and the distribution of an angular distribution in the category without a high-\PT jet.

This paper is organized as follows: a brief description of the CMS detector is given in Section~\ref{sec:detector}. The simulated samples used in this analysis are discussed in Section~\ref{sec:samples}, followed by an overview of the event and object reconstruction in Section~\ref{sec:objects}.
In Section~\ref{sec:selections}, the event selection and the observables for signal extraction are described, after which the background estimation is described in Section~\ref{sec:background}.
The systematic uncertainties are discussed in Section~\ref{sec:systematics}. The results and their statistical interpretation are presented in Section~\ref{sec:results}. The paper closes with a summary in Section~\ref{sec:summary}.

Tabulated results are provided in the HEPData record for this analysis~\cite{hepdata}.

\section{The CMS detector} \label{sec:detector}

The central feature of the CMS apparatus is a superconducting solenoid of 6\unit{m} internal diameter, providing a magnetic field of 3.8\unit{T}. Within the solenoid volume are a silicon pixel and strip tracker, a lead tungstate crystal electromagnetic calorimeter (ECAL), and a brass and scintillator hadron calorimeter (HCAL), each composed of a barrel and two endcap sections. Forward calorimeters extend the pseudorapidity ($\eta$) coverage provided by the barrel and endcap detectors. Muons are measured in gas-ionization detectors embedded in the steel flux-return yoke outside the solenoid. A more detailed description of the CMS detector, together with a definition of the coordinate system used and the relevant kinematic variables, can be found in Ref.~\cite{CMS}.

Events of interest are selected using a two-tiered trigger system~\cite{CMS:2016ngn}. The first level, composed of custom hardware processors, uses information from the calorimeters and muon detectors to select events at a rate of around 100\unit{kHz} within a time interval of about 4\mus~\cite{CMS:2020cmk}. The second level, known as the high-level trigger, consists of a farm of processors running a version of the full event reconstruction software optimized for fast processing, and reduces the event rate to about 1\unit{kHz} before data storage~\cite{CMS:2016ngn}.

During the 2016 and 2017 data taking, a gradual shift in the timing of the inputs of the ECAL first-level trigger in the region at $\abs{\eta} > 2.0$ caused a specific trigger inefficiency. For events containing an electron (a jet) with transverse momentum \pt larger than $\approx$50\GeV ($\approx$100\GeV), in the region $2.5 < \abs{\eta} < 3.0$ the efficiency loss is $\approx$10--20\%, depending on \pt, $\eta$, and time. Correction factors were computed from data and applied to the kinematic acceptance evaluated by simulation.

\section{Simulated samples} \label{sec:samples}

Samples of simulated events are used to devise selection criteria, and estimate and validate background predictions.
The main sources of background are the pair production of \PQt quarks (\ttbar), single \PQt quark production, Drell--Yan (DY) and \PW boson production in association with jets, respectively denoted as ``\DYjets'' and ``\Wjets'', diboson ($\PW\PW$, $\PW\PZ$, $\PZ\PZ$) production, and quantum chromodynamics (QCD) production of multijet events.
The backgrounds are simulated with the Monte Carlo technique, except for the QCD multijet background in the \mumu and \emu channels, and the background due to jets misidentified as \tauh in the \mutau, \etau, and \tautau channel.
The \Wjets and \DYjets processes are generated using the \MGvATNLO~\cite{aMCatNLO} generator~(versions 2.2.2 and 2.3.3) at LO precision with up to four additional partons simulated at the matrix element level, which are merged and matched to the parton shower with the MLM scheme~\cite{MLM}. 
Diboson production is generated at LO with \PYTHIA~\cite{Pythia8_2015} in version 8.212 (8.230) for the 2016 (2017 and 2018) sample, except for the gluon-initiated process, which is simulated with the \MCFM (version 7.0) generator~\cite{Campbell:1999ah,sigma_NLO_diboson,Campbell:2015qma}.
The \ttbar and single \PQt quark production are generated by \POWHEG~\cite{POWHEG0,POWHEG1,POWHEG2} (versions 2.0 and 1.0) at next-to-LO (NLO) precision in perturbative QCD~\cite{POWHEG_TT,POWHEG_TT2,POWHEG_ST_st,POWHEG_ST_tW}.

{\tolerance=800 Minor background processes, denoted in figures and tables as ``other'', include electroweak production of \PW or \PZ bosons in association with jets, the associated production of a \PW or \PZ boson, a photon, and additional jets (\Vgjets), as well as the production of three massive gauge bosons (triboson production). All minor backgrounds are simulated using different versions of \MGvATNLO depending on the process and data-taking period. The production of a \PZ boson and a photon, as well as that of three massive gauge bosons, is simulated at NLO, while all other minor backgrounds are generated at LO.\par}

The \DYjets, \Wjets, \ttbar, and single \PQt processes are normalized using cross sections computed at next-to-NLO (NNLO) in perturbative QCD~\cite{sigma_DY,sigma_TT,sigma_ST}. All other background processes are normalized according to their respective cross sections computed at the same order as that of the matrix element calculations.

A simplified $\widetilde{R}_2$ model~\cite{RD_LQ_pheno9} is used to generate the signal samples with a scalar \LQ.
In the case of the vector \LQ, a $U_1$ model~\cite{LQ_theory_NLO} is used, in which the \LQ couples only to left-handed fermions of the third generation, with a corresponding Lagrangian density
$\mathcal{L} \supset \lambda\big(\PAQb_\mathrm{L}\gamma^\mu\PGt_\mathrm{L}U_{\mu}+\PAQt_\mathrm{L}\gamma^\mu\PGn_\mathrm{\PGt L}U_{\mu}\big)+\text{h.c.}$. The coupling of \LQs to quarks and charged leptons relative to the coupling to quarks and neutrinos is expressed by a dimensionless parameter $\beta$. For all signal samples we consider exclusive \LQ couplings to \PQb quarks and \PGt leptons, corresponding to $\beta=1$.
The signal samples are generated at LO precision using the \MGvATNLO 2.6.1~\cite{MadGraph} event generator with the five-flavor scheme.
Since the kinematic properties of the final state are expected to be independent of $\lambda$ for the \LQ signals of $t$-channel exchange, pair production, and single production with approximately $\lambda < 1.5$, and the \LQ decay width is smaller than the experimental resolution, samples generated with $\lambda=1$ are used for those cases. Additional samples of single \LQ production are generated for values of $\lambda$ between 1.5--2.5 to take into account the increased width and changes to kinematic properties of the final state.
The single and pair production of vector \LQs depends on an additional parameter $\kappa$, which affects the interaction of the \LQ with the SM gauge fields. Values of $\kappa = 1$ and 0 are considered in this analysis, corresponding to the Yang--Mills and minimal coupling scenarios, respectively.
The nonresonant production of two \PGt leptons via $t$-channel \LQ exchange is generated assuming no interference with the SM DY process, as the corresponding reduction of the signal yield in sensitive regions of the analysis would be less than 5\% for $\lambda>1$~\cite{LQ_large_coupling} and is therefore considered to be negligible. 
The cross sections of all signal processes are computed at LO with the models discussed above, except for the production of a scalar \LQ pair, which is computed at NLO~\cite{Kramer:1997hh,Kramer:2004df,LQ_scalar_NLO-NNLL}.

The event generators are interfaced with \PYTHIA 8.212 for 2016 and 8.230 for 2017 and 2018 samples to model the parton showering and fragmentation, as well as the decay of the \PGt leptons. The \PYTHIA parameters affecting the description of the underlying event are set to the CUETP8M1 (CP5) tune for all 2016 (2017--2018) samples~\cite{TuneCUETP8M1,TuneCP}, except for the 2016 \ttbar sample, for which CUETP8M2T4~\cite{TuneCUETP8M2T4} is used. For the simulation of the SM background processes, the NNPDF3.0 PDFs~\cite{NNPDF30} with the order matching that of the matrix element calculations are used with all generators for 2016, while NNPDF3.1~\cite{NNPDF31} NNLO PDFs are used for 2017 and 2018 running conditions. For the simulation of the \LQ signal, NNPDF3.1 PDFs at LO are used for all years.
Generated events are processed through a simulation of the CMS detector based on \GEANTfour~\cite{GEANT4}, and are reconstructed with the same algorithms as used for data.
The simulated samples include additional \pp interactions per bunch crossing, referred to as ``pileup''.
The effect of pileup is taken into account by generating minimum-bias collision events with \PYTHIA. The simulated events are weighted such that the distribution of the number of pileup interactions matches that in data.

\section{Event and object reconstruction} \label{sec:objects}

A particle-flow algorithm~\cite{PF2017} aims to reconstruct and identify each individual particle in an event, with an optimized combination of information from the various elements of the CMS detector. Photons are identified as ECAL energy clusters not linked to the extrapolation of any charged particle trajectory to the ECAL. Electrons are identified as a primary charged particle track and potentially many ECAL energy clusters, corresponding to this track extrapolation to the ECAL and to possible bremsstrahlung photons emitted along the way through the tracker material. Muons are identified as tracks in the central tracker consistent with either a track or several hits in the muon system, and associated with calorimeter deposits compatible with the muon hypothesis. Charged hadrons are identified as charged particle tracks neither identified as electrons, nor as muons. Finally, neutral hadrons are identified as HCAL energy clusters not linked to any charged hadron trajectory, or as a combined ECAL and HCAL energy excess with respect to the expected charged hadron energy deposit.

The energy of photons is obtained from the ECAL measurement. The energy of electrons is determined from a combination of the track momentum at the main interaction vertex, the corresponding ECAL cluster energy, and the energy sum of all bremsstrahlung photons attached to the track. The energy of muons is obtained from the corresponding track momentum. The energy of charged hadrons is determined from a combination of the track momentum and the corresponding ECAL and HCAL energies, corrected for the response function of the calorimeters to hadronic showers. Finally, the energy of neutral hadrons is obtained from the corresponding corrected ECAL and HCAL energies. 

The primary \pp interaction vertex of an event is taken to be the reconstructed vertex corresponding to the hardest scattering in the event, evaluated using tracking information alone, as described in Section 9.4.1 of Ref.~\cite{CMS_vertex_phase2}.

The momentum resolution for electrons with a $\pt \approx 45\GeV$ from \Zee decays ranges from 1.6--5.0\%. It is generally better in the barrel region than in the endcaps, and also depends on the bremsstrahlung energy emitted by the electron as it traverses the material in front of the ECAL~\cite{CMS-DP-2020-021,CMS:2020uim}. 
Electrons are identified with a multivariate analysis~\cite{TMVA} discriminant that includes several quantities describing the track quality, the shape of the energy deposits in the ECAL, and the compatibility of the measurements from the tracker and the ECAL.
Electrons must also pass a filter that rejects electrons coming from photon conversions and are required to be isolated~\cite{CMS:2020uim}. Electrons are required to have $\pt>50\GeV$, $\abs{\eta}<2.3$, and be within a longitudinal distance $d_z$ of 0.2\cm and a radial distance $d_{xy}$ of 0.045\cm from the primary vertex (PV).

Matching muons to tracks measured in the silicon tracker results in a relative \pt resolution, for muons with \pt up to 100\GeV, of 1\% in the barrel and 3\% in the endcaps. The \pt resolution in the barrel is better than 7\% for muons with \pt up to 1\TeV~\cite{CMS_muon_2018}. 
Muons are identified with requirements on the quality of the track reconstruction and on the number of measurements in the tracker and the muon system, and are required to be isolated~\cite{CMS_muon_2018}. Furthermore, muons must have $\pt>50\GeV$, $\abs{\eta}<2.4$, and be within a longitudinal distance $d_z<0.2\cm$ and a radial distance $d_{xy}<0.045\cm$ from the PV.

For each event, hadronic jets are clustered from reconstructed particles using the infrared- and collinear-safe anti-\kt algorithm~\cite{antikT,fastjet2} with a distance parameter of 0.4. Jet momentum is determined as the vectorial sum of all particle momenta in the jet, and is found from simulation to be, on average, within 5--10\% of the true momentum over the entire \pt spectrum and detector acceptance. Pileup interactions can contribute additional tracks and calorimetric energy depositions to the jet momentum. To mitigate this effect, charged particles identified to be originating from pileup vertices are discarded and an offset correction is applied to correct for remaining contributions. Jet energy corrections are derived from simulation to bring the measured response of jets to that of particle level jets on average. In situ measurements of the momentum balance in dijet, $\text{photon} + \text{jet}$, $\PZ + \text{jet}$, and multijet events are used to account for any residual differences in the jet energy scale between data and simulation~\cite{CMS-JME-13-004}. The jet energy resolution amounts typically to 15--20\% at 30\GeV, 10\% at 100\GeV, and 5\% at 1\TeV~\cite{CMS-JME-13-004}. Additional selection criteria are applied to each jet to remove jets potentially dominated by anomalous contributions from various subdetector components or reconstruction failures. In this analysis, jets are required to have $\pt>50\GeV$ and $\abs{\eta}<4.7$, and must be separated from the selected leptons by $\DR = \sqrt{\smash[b]{(\Delta\eta)^2 + (\Delta\phi)^2}} > 0.5$, where $\phi$ denotes the azimuthal angle.

Jets originating from the hadronization of \PQb quarks are identified using the \textsc{DeepCSV} algorithm~\cite{btag_deepCSV, btag_2018} based on a deep neural network with four hidden layers that is an extension of the combined secondary vertex algorithm~\cite{btag_2013}. It exploits observables related to the long lifetime and large mass of \PQb hadrons. This analysis uses a fixed working point (WP) of \textsc{DeepCSV} that corresponds to a 10\% misidentification rate for jets originating from light quarks or gluons, and an efficiency of up to 90\% in selecting \PQb quark jets, depending on the jet \pt and $\eta$. 
The efficiency degrades to approximately 60\% for $\pt > 500\GeV$. In this analysis, \PQb-tagged jets are required to have $\pt>50\GeV$ and $\abs{\eta} < 2.4$ (2.5) for the 2016 (2017--2018) data-taking period.

{\tolerance=800 Hadronic \PGt lepton decays are reconstructed from jets, using the hadrons-plus-strips algorithm~\cite{HPS2}, which combines 1 or 3 tracks with energy deposits in the calorimeters, to identify the \PGt lepton decay modes. Neutral pions are reconstructed as strips with dynamic size in $\eta$-$\phi$ from reconstructed electrons and photons, where the strip size varies as a function of the \pt of the electron or photon candidate. To distinguish genuine \tauh decays from jets originating from the hadronization of quarks or gluons, electrons, or muons, the \DeepTau algorithm~\cite{DeepTau} is used, which combines information from all individual reconstructed particles near the \tauh candidate axis with properties of the \tauh candidate and the event. The probability of a jet to be misidentified as a \tauh candidate by the \DeepTau algorithm depends on the \pt and quark flavor of the jet. In simulated \Wjets events it has been estimated to be 0.43\% for a genuine \tauh identification efficiency of 70\%. The misidentification rate for electrons (muons) is 2.60\,(0.03)\% for a genuine \tauh identification efficiency of 80 ($>$99)\%. Lastly, the \tauh candidate is required to have $\pt>50\GeV$, $\abs{\eta}<2.3$, and be within a longitudinal distance $d_z$ of 0.2\cm from the PV.\par}

The missing transverse momentum vector \ptvecmiss is computed as the negative vector \pt sum of all the particle-flow candidates in an event, and its magnitude is denoted as \ptmiss~\cite{CMS:2019ctu}. The \ptvecmiss is modified to account for corrections to the energy scale of the reconstructed jets in the event.

\section{Event selection} \label{sec:selections}

This search targets a final state signature with energetic \PGt leptons and \PQb quark jets. 
Section~\ref{sec:baseline} introduces a baseline event selection, which requires the events to be consistent with either the \etau, \mutau, \ditau, \emu, or \mumu decay channel, where the electron, muon, or \tauh object is assumed to be the visible product of \PGt lepton decays.
The most sensitive decay channels are \etau, \mutau, and \ditau, while the expected \LQ signal contribution is negligible compared to the background in the \emu and \mumu channels. However, the fully leptonic channels still provide a large sample of background events that help to constrain the systematic uncertainties in the simulated background in a combined fit.
Event categories that target the resonant \LQ production based on events with at least one high-\pt jet not originating from a \tauh decay are then defined in Section~\ref{sec:selections_res}, and those targeting a nonresonant signal based on events with no such high-\pt jets are defined in Section~\ref{sec:selections_nonres}.

\subsection{Baseline selection} \label{sec:baseline}
This analysis makes use of three types of triggers depending on the explored final state. Events collected by single-electron triggers are considered in the \etau channel, by single-muon triggers for the \mutau, \emu, and \mumu final states, and by triggers requiring a pair of \tauh candidates for the \ditau channel. A minimum \pt of 27--35\GeV (24--27\GeV, 35--40\GeV) is required for each particle candidate to pass the single-electron (single-muon, \ditau) triggers, varying between years of data taking. The \ditau triggers are limited to $\abs{\eta}<2.1$, therefore \tauh candidates are required to have $\abs{\eta}<2.1$ in the \ditau channel.

The \etau (\mutau) channel requires exactly one electron (muon) and at least one \tauh candidate passing the reconstruction criteria described in Section~\ref{sec:objects}.
The selected lepton and \tauh must have an angular separation of $\DR > 0.5$ and have opposite-sign (OS) electric charge. If multiple such combinations of a lepton and a \tauh candidate are found, the pair with the highest-\pt \tauh candidate is selected.
In order to reject background events from \Zeepm and \Zmmpm decays, the \DeepTau discriminators against leptons misidentified as \tauh candidates are applied to the \tauh object in both channels. In the \etau (\mutau) channel, the ``Loose'' (``VVLoose'') WP of the discriminator against electrons, as defined in Ref.~\cite{DeepTau}, is used. In both channels, the ''VLoose'' WP of the discriminator against muons is employed.

Lastly, for both the \etau and \mutau channels, a veto on additional leptons is applied in order to reduce \DYjets, \ttbar, and diboson backgrounds, as well as to keep orthogonality between the \etau and \mutau channels. For this purpose, events with additional isolated electrons or muons that pass looser identification criteria are discarded.

Events in the \ditau channel are required to have two \tauh candidates satisfying the criteria detailed in Section~\ref{sec:objects}. The two \tauh candidates must be separated by $\DR > 0.5$ and have OS electric charges as well.
As in the case of the semileptonic channels, extra-lepton vetoes, and the \DeepTau discriminators against leptons are applied. As there is little background in this channel from electrons or muons misidentified as \tauh candidates, loose WPs of the discriminators against electrons (``VVLoose'') and muons (``VLoose'')  are used.

In the \emu (\mumu) channel, the presence of exactly one electron and one muon (exactly two muons) of OS electric charges and with $\DR > 0.5$ is required, while a veto is applied on additional leptons. The criteria discussed in Section~\ref{sec:objects} are applied.

\subsection{Search for resonant LQ production} \label{sec:selections_res}
The final states of the single- and pair-production processes are characterized by two high-\pt \PGt leptons and one or two high-\pt \PQb quarks. 
In order to obtain the highest sensitivity to the \LQ signal, the event categorization and selections have been optimized. 
The kinematic variable showing the best separation power between signal and background is the scalar sum \STMET of the \pt of the \PGt decay candidates ($\pt^1$, $\pt^2$), the leading jet ($\pt^{\text{j}}$), and \ptmiss,
\begin{linenomath}
\begin{equation} \label{eq:ST_MET}
  \STMET = \pt^1 + \pt^2 + \pt^{\text{j}} + \ptmiss.
\end{equation}
\end{linenomath}
The SM background has a steeply falling \STMET distribution, while the \LQ signal is expected to appear at large values of \STMET.
The signal strength is extracted from a simultaneous fit to the observed distribution of this observable.
This analysis considers events that have at least one jet with $\PT>50\GeV$. Two orthogonal event categories are then constructed: the ``0b'' category, which requires zero loosely \PQb-tagged jets with $\pt>50\GeV$, and the ``\geqb'' category, which requires at least one such jet. The 0b category primarily targets events of single or pair production of \LQs in which jets initiated by a \PQb quark are not \PQb tagged.
An additional requirement is applied on the invariant mass \mvis of the sum of the four-momenta of all visible decay products of the two \PGt leptons, $\mvis > 100\GeV$, to remove DY events with negligible loss of signal.

\begin{figure}[bht]
  \centering
  \includegraphics[width=\cmsFigWidth]{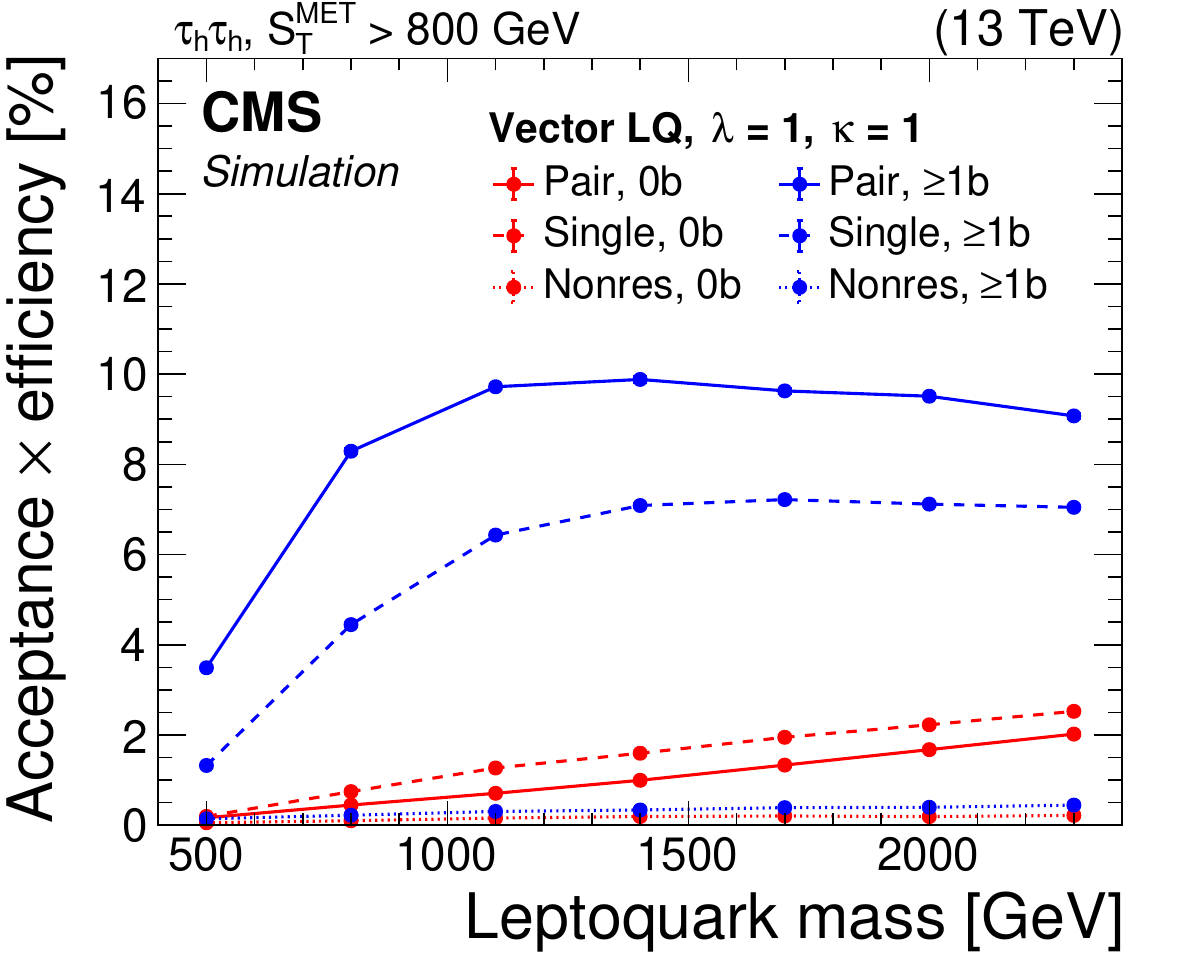}
  \includegraphics[width=\cmsFigWidth]{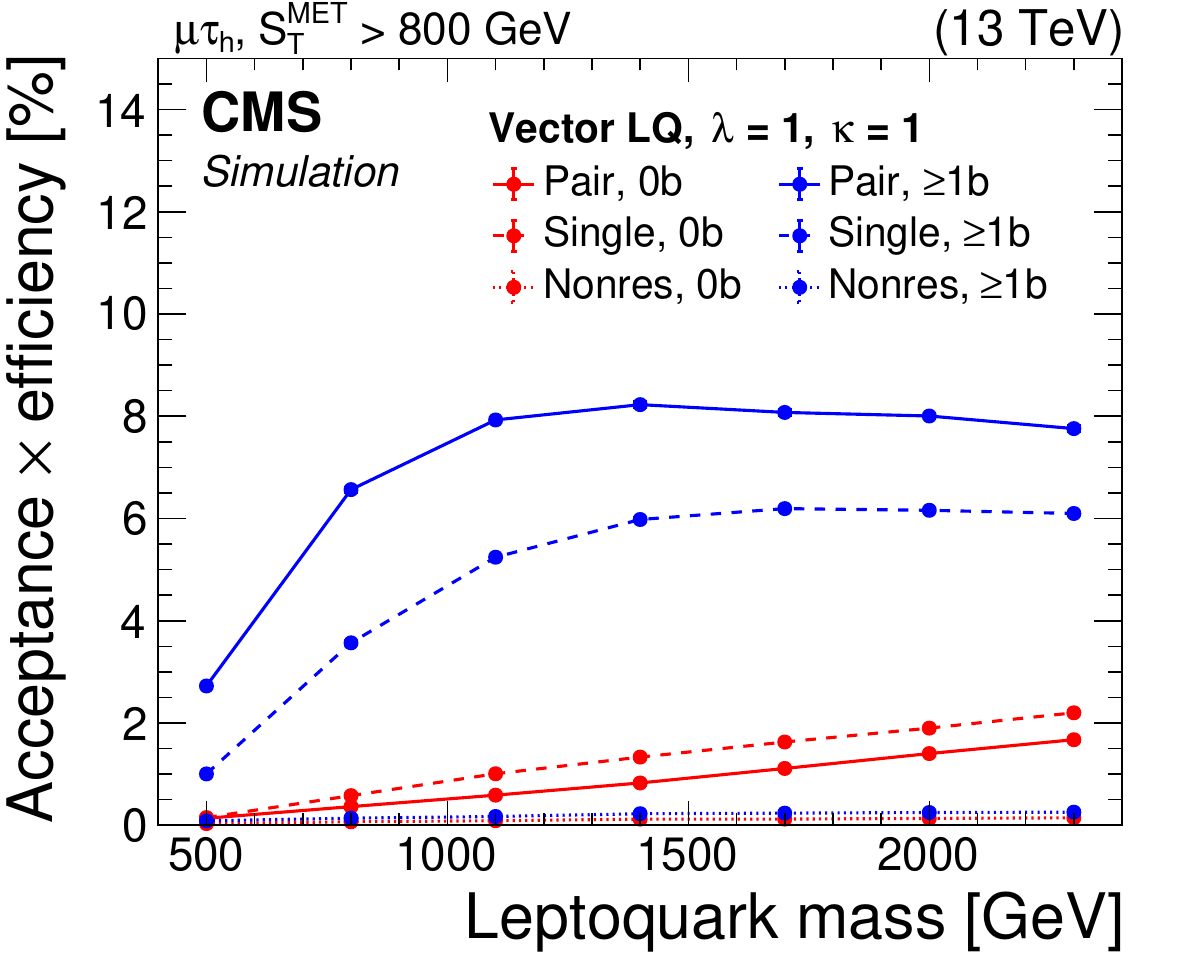}
  \includegraphics[width=\cmsFigWidth]{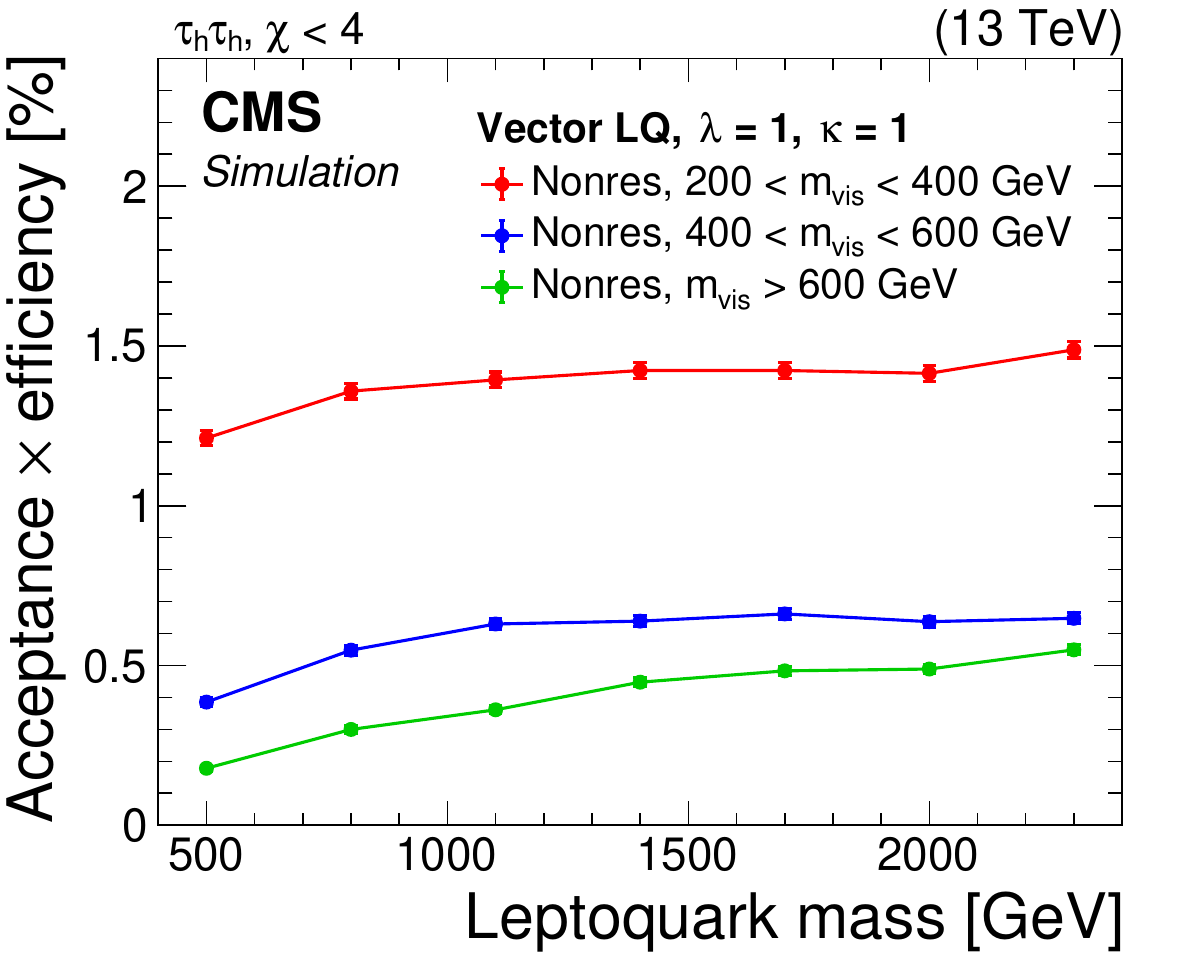}
  \includegraphics[width=\cmsFigWidth]{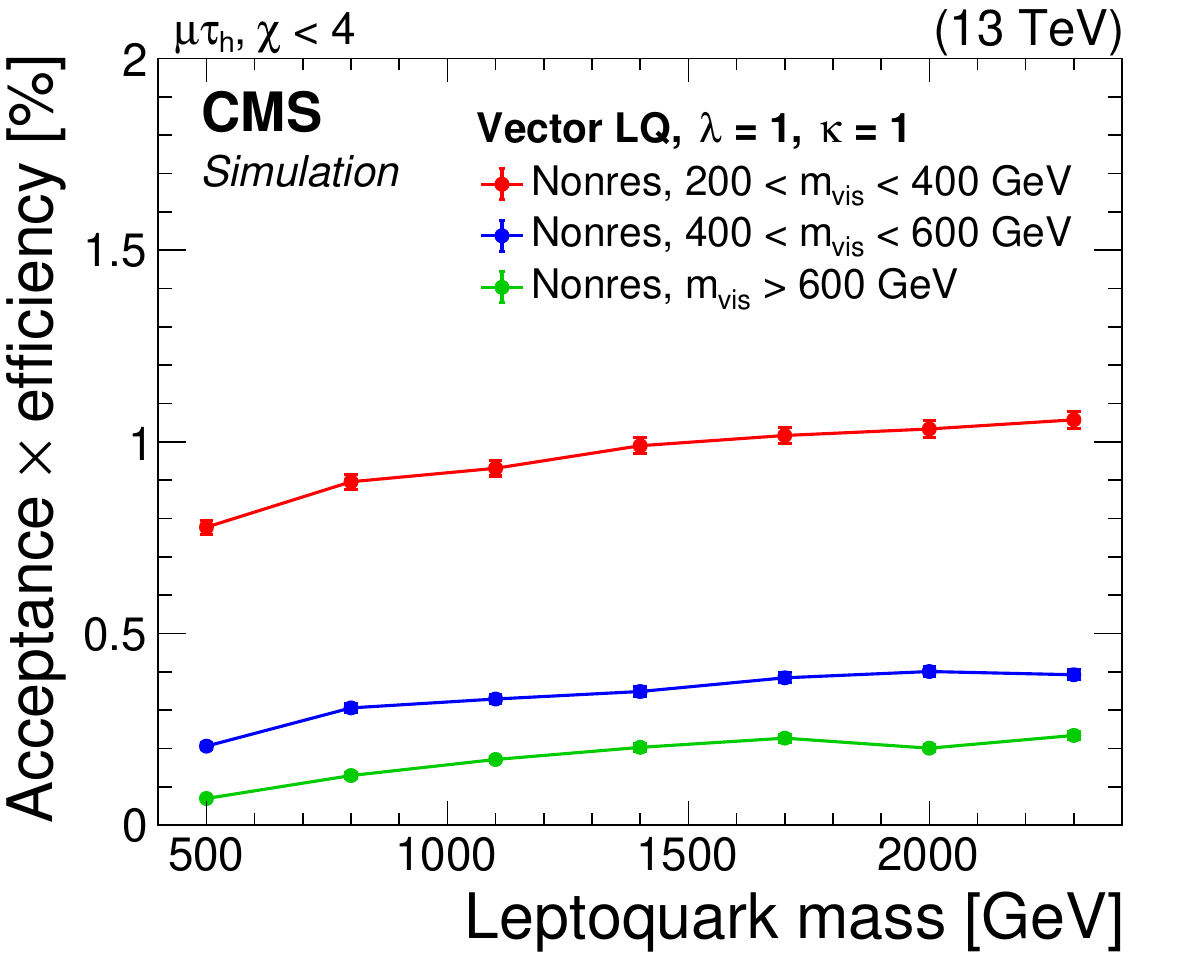}
  \caption{
Product of acceptance and efficiency for a vector \LQ signal in the \ditau (left) and \mutau (right) channels of the 0b and \geqb (upper), and the 0j categories (lower).
The resonant \LQ signals are neglected in the 0j category.
Vertical bars (only just visible) indicate the statistical uncertainty.
  }
  \label{fig:acc_eff}
\end{figure}

The product of acceptance and efficiency for the vector \LQ signal in the resonant signal categories 0b and \geqb is shown in the upper row of Fig.~\ref{fig:acc_eff} as a function of \mLQ. The scalar \LQ signal shows a very similar behavior.
The acceptances and efficiencies are restricted to the sensitive region of $\STMET > 800\GeV$ and are computed with respect to all possible decay modes of the two \PGt leptons.
In the \ditau channel of the \geqb category, the product of acceptance and selection efficiency for the \LQ pair production events varies with \mLQ between approximately 3.5 and 10.0\%, while it varies between 1.0 and 7.5\% for the single production.
Below \LQ masses of approximately 1\TeV, the acceptance and efficiency for resonant production decrease towards lower masses in the \geqb category because of a softer \pt spectrum.
Contributions from the nonresonant signal are also considered in the 0b and \geqb categories, as one or more high-\PT jets due to initial-state radiation or gluon splitting may be present.

\subsection{Search for nonresonant LQ production} \label{sec:selections_nonres}
To increase the purity of the nonresonant signal, a small longitudinal Lorentz boost $\abs{\eta_{1}+\eta_{2}}/2<1.1$ and a pseudorapidity difference $\abs{\Delta\eta} <3$ between the two \PGt decay candidates are required. To make this category orthogonal to those described in Section~\ref{sec:selections_res}, selected events are required to have no jets with $\pt > 50\GeV$. This category will therefore be referred to as the ``0j'' category.
After these selections, the events are separated into three bins of the visible mass \mvis: 200--400, 400--600, and larger than 600\GeV, because the signal purity increases with \mvis.

In these three categories, the sensitive observable used for the signal extraction in a simultaneous fit is the angular separation between the two \PGt lepton candidates, defined as $\chi = \exp\left(\abs{\Delta\eta}\right)$.
The choice of this variable is motivated by the fact that the $\chi$ distribution in Rutherford scattering is flat~\cite{Combridge:1983jn},
while new physics processes with a more isotropic angular distribution will appear as an excess at low $\chi$ values.
The $\chi$ distribution is then measured over the range $1 < \chi < 21$, which implies a maximum value of $\abs{\Delta\eta} = 3.04$.

The product of acceptance and efficiency for the vector \LQ signal in the 0j categories is shown in the lower row of Fig.~\ref{fig:acc_eff} as a function of \mLQ. Very similar behavior is found for the scalar \LQ signal.
The acceptances and efficiencies are restricted to the sensitive region of $\chi < 4$ and are computed with respect to all possible decay modes of two \PGt leptons.

\section{Background estimation} \label{sec:background}

For the search of a resonant \LQ signal in the 0b and \geqb categories, the dominant backgrounds in most search channels stem from \ttbar production because of the presence of genuine electrons, muons, \PGt leptons, and \PQb quark jets from \PQt quark decays. In final states with at least one \tauh candidate, events with jets misidentified as \tauh candidates (\jtf events) are another major source of background. For the nonresonant search in the 0j category, the main backgrounds in the signal region (SR) are DY production as well as the \jtf events.

The shape of the \PZ boson \pt spectrum of simulated \DYjets events is adjusted by simulation-to-data correction factors. These factors are derived as a function of the \pt of the selected dimuon system in a sample enriched with \DYjets dimuon events. This procedure is referred to as ``\PZ boson \pt reweighting'' in the following.

\subsection{Background estimation for the \texorpdfstring{\ltau}{ltauh} and \texorpdfstring{\ditau}{tauhtauh} channels}
For the \etau, \mutau, and \ditau channels, all SM background processes are estimated using simulation, except for background of \jtf events. The \jtf background in the different signal channels consists of QCD multijet, \Wjets, and \ttbar events. This background is estimated with the so-called ``fake-factor'' (FF) method that is described in detail in Refs.~\cite{FF_MSSM,FF_ZTT}. Depending on the event category, the \jtf background constitutes approximately 25--71\% of the total background in the \etau and \mutau channels and 46--82\% in the \ditau channel, decreasing with the number of b-tagged jets and the minimum requirement imposed on \mvis.

The FF method proceeds by defining an application region (AR) containing the same selection criteria as in the SR, with the exception that the requirement on the \DeepTau discriminator against jets is inverted for one \tauh candidate, while still requiring it to pass the ``VLoose'' WP.
The AR is primarily populated by events with jets misidentified as \tauh candidates, and has contamination from genuine \tauh decays at the level of a few percent or below.
The FF represents the ratio of the number of events with a jet misidentified as a \tauh candidate in the AR to that in the SR. For both the AR and SR, corresponding control regions (CRs) are defined, in which the same requirements on the \DeepTau discriminator against jets as in the AR and SR are applied. The FF is measured in data as the ratio of the number of events in each of the two CRs, in which a jet was misidentified as a \tauh candidate.

The FF is measured separately for the three dominant background processes using dedicated CRs that are pure in the respective background in order to take into account the differing compositions of jet flavors. The FFs for QCD multijet events are estimated in a CR with same-sign (SS) \tauh pairs, while the FFs for \Wjets events are estimated in \ltau final states with no \PQb-tagged jets and a high transverse mass \mT of the lepton and \ptvecmiss. The transverse mass is defined as $\mT = \sqrt{\smash[b]{2 \pt^\ell \ptmiss \left(1-\cos\Delta\phi\right)}}$, where $\Delta\phi$ denotes the azimuthal distance between the lepton and \ptvecmiss. For \ttbar events, the FF are estimated purely from simulated samples. The FFs measured separately for the three background processes are combined in a weighted sum, in which each FF is weighted with the fraction of events of the corresponding process in the AR. The \jtf background in the SR is obtained by applying the combined FF to the data in the AR, from which simulated events with genuine \tauh candidates have been subtracted. The combined FF takes an average value between 0.40 and 0.47, depending on the category.

The FF is calculated as a function of the \pt of the \tauh candidate and the number of jets in the event. In \ltau events, the FF is also measured as a function of the {\DR} between the \tauh candidate and the light lepton. A polynomial function is fitted to the FF values as a function of \pt to interpolate between measured values. The presence of small backgrounds in the AR that contain genuine \tauh decays results in an underestimation of the number of events with a misidentified \tauh in the SR of up to 2\%. Therefore, a correction is applied based on the fractions of these processes in simulated events. Corresponding uncertainties are incorporated into the fit model of the SR, as described in Section~\ref{sec:systematics}.
The FF method for the \ditau channel is similar, with the exception that a weight factor of 0.5 is applied to take into account combinatorial effects due to the presence of two \tauh candidates.

The FF method is tested in two validation regions (VRs): one is constructed by inverting the leading jet \pt requirement (\ie, $\pt < 50\GeV$) and the other by using events with two \tauh candidates that do not fulfill the \DeepTau identification requirement against jets used to define the SR.
For both VRs, all other selection criteria are kept identical to the SR, except that the \PQb tagging requirement is removed to increase the number of events. The signal contamination is negligible in both VRs. Data and the predicted background are found to agree within statistical uncertainties, demonstrating the validity of the FF method.

\subsection{Background estimation for the \texorpdfstring{\emu}{emu} and \texorpdfstring{\mumu}{mumu} channels} \label{sec:background_emu}
In the \emu and \mumu channels, all SM background processes, except for the QCD multijet background, are estimated using simulation and normalized according to the theoretical cross sections.

The QCD multijet background is small and is estimated from an SS CR. In this CR, the QCD multijet background shape is constructed by subtracting the simulated SM backgrounds from observed data in each histogram bin. This shape is scaled by an extrapolation factor to obtain the expected QCD multijet background in the OS SR. The extrapolation factor is measured in simulated QCD multijet events by inverting the lepton isolation requirement, and is found to be approximately 2.4.

\section{Systematic uncertainties} \label{sec:systematics}

Various systematic uncertainties are taken into account as nuisance parameters in the final fit of the expected background and signal to the data. Uncertainties are treated differently depending on whether they affect only the rate of a given process or both its rate and shape in the distributions used in the final fit.

\subsection{Rate uncertainties}
The uncertainty in the integrated luminosity varies depending on the year~\cite{CMS-LUM-17-003,CMS-PAS-LUM-17-004,CMS-PAS-LUM-18-002}, amounting to 1.6\% for the 2016--2018 sample, and affects the rate of the signal and background processes that are based on simulation.
Uncertainties in the electron or muon identification and trigger efficiency amount to 2\% each~\cite{HTT_2016_paper}. For events in which electrons or muons are misidentified as \tauh candidates, predominantly \Zeepm events in the \etau channel and \Zmmpm events in the \mutau channel, rate uncertainties of 12 and 25\%~\cite{HPS3}, respectively, are applied, as determined via the ``tag-and-probe'' method~\cite{CMS:2011aa}.

The QCD multijet background estimation in the \emu and \mumu channels is found to have rate uncertainties up to 20\%.
The uncertainty in the \DYjets cross section in the \geqb category is estimated using a dedicated CR of events with two \tauh candidates and at least one \PQb-tagged jet and $\mvis<100\GeV$. A 20\% uncertainty is assigned to the \DYjets cross section in the \geqb category based on the total number of observed data events compared to the expected background in this CR. In the other signal categories, the uncertainty in the NNLO computation is given as 3\%. The uncertainties in the cross section for the \ttbar, diboson, and single \PQt quark processes are 5.5, 6, and 5.5\%, respectively. A value of 5, 10, and 25\% is assigned to the uncertainty in the cross section for \Vgjets, electroweak \PW or \PZ boson, and triboson production, respectively. These uncertainties are correlated between the years but assumed to be uncorrelated between 0j and the other signal categories.

\subsection{Shape uncertainties} \label{sec:systematics_shapes}
Differences in the \tauh identification efficiency between data and simulation are corrected by scale factors (SFs) that are applied as functions of the \tauh decay mode for the \ditau channel. The uncertainty in these SFs is predominantly statistical and assumed to be uncorrelated between the three data-taking years and the four \tauh decay modes considered. In the \mutau and \etau channels instead, an inclusive SF is applied for each year, and its uncertainty is assumed as uncorrelated between the three years but fully correlated between both channels. An additional uncertainty, fully correlated between all analysis regions, is assigned to account for the extrapolation of each \tauh identification SF from $\pt=40\GeV$ to higher values, covering a potential \pt-dependence of the SFs. The uncertainty is expanded linearly up to values of $+$8 and $-$17\% at $\pt=500\GeV$, beyond which it is continued as a constant. The uncertainty in the trigger efficiency SF in the \ditau channel is taken into account by varying the SF within its total uncertainty. This uncertainty is considered uncorrelated across years and fully correlated between event categories in the \ditau channel.

The uncertainty in the \tauh energy scale depends on the \tauh decay mode and \pt, and varies between 0.6--1.4\% for $\pt<34\GeV$, and between 1--4\% for $\pt>170\GeV$, with a linear interpolation between the respective uncertainties for $34<\pt<170\GeV$.
It is only applied to the simulated events with genuine \tauh candidates, namely those in \DYjets, \ttbar, and \LQ events. It is uncorrelated between the years but correlated between the different channels.
For events in which a muon is misidentified as a \tauh candidate, a shape uncertainty is derived by varying the \tauh energy scale by 1\%. For events with electrons that are misidentified as a \tauh candidate, the uncertainties in the measured energy corrections are used, which vary from 0.8 to 6.6\% depending on the \tauh decay mode and pseudorapidity region.

The uncertainties in the jet energy scale are up to 4\%, depending on the values of \pt and $\eta$ of the jet~\cite{CMS-JME-10-011}. The uncertainties in the jet energy scale and resolution are taken into account as shape uncertainties by taking the one-standard-deviation variations of the corrections and propagating them to all the affected variables in the analysis (including jet \pt, jet multiplicity, \PQb tag multiplicity, \ptmiss and \STMET).
These uncertainties are assumed to be uncorrelated between the years and correlated between the three signal categories and different \PGt decay channels. The uncertainty in the correction of the ECAL timing shift at the trigger level is taken into account by varying the correction factor within its uncertainties.

The uncertainties in the \PQb tagging efficiency of heavy-flavor jets (\PQb and \PQc quark jets) and the misidentification rate of light-flavor (\PQu, \PQd, \PQs quarks) and gluon jets are propagated from the uncertainties in the \PQb tagging SFs. The efficiencies for \PQb and \PQc quarks are assumed to be correlated, as are the misidentification rates for light-flavor and gluon jets.
Most of the shape variations have little dependence on \STMET, with about a 1\% change in overall yield in the \geqb category, and less than 3\% in 0b. The uncertainties are partially correlated between the years.

The uncertainties related to the FF method include the total systematic uncertainty in the measurement of the FFs in each of the \Wjets, \ttbar, and QCD CRs, as well as the statistical uncertainty in the \pt-interpolation of the FFs.
They are treated as uncorrelated between the years, event categories, and \tautau final states, as the dominant background process can differ and the FF measurement is dominated by the statistical uncertainty due to the limited number of events in the CRs.
In addition, the statistical uncertainty in the FF interpolation is considered uncorrelated between four bins in \tauh \pt ($<$100, 100--200, 200--500, $>$500\GeV) to allow independent variations of the FFs in regions with different kinematic properties.

The uncertainty in the \PZ boson \pt reweighting is conservatively estimated by varying the simulation-to-data correction factors up and down by 50\%. Similarly, an uncertainty is assigned to the modeling of the \PQt quark \pt in simulated \ttbar, which was found to differ between data and simulation~\cite{CMS:2021vhb, ATLAS:2022xfj, ATLAS:2022mlu}.
Distributions of \STMET and $\chi$ with differing shapes are created by applying corrections to the simulated \PQt quark pt distribution of the form described in Section 3 of Ref.~\cite{CMS:2021iuw} but with a much larger range of variations in order to ensure that the final fit is only weakly constrained by this uncertainty.
Owing to large pre-fit uncertainties, the \PZ boson and \PQt quark \pt distributions are modeled using data in the \emu and \mumu regions in the final maximum likelihood fit.

The renormalization and factorization scales \muR and \muF used in the event generation depend on the process simulated and the generator used. They are either set to fixed values or are related to kinematic properties of the individual generated events. The shape uncertainties due to these scales are obtained by varying \muR and \muF simultaneously by the same or a different factor (0.5, 1, or 2) while keeping the total yield constant and taking the envelope of all variations. Extreme combinations of $(\muR,\,\muF)$ varied by factors of $(0.5,\,2)$ or $(2,\,0.5)$ are excluded. Similarly, the shape uncertainties due to the PDFs are taken into account by creating a symmetric envelope~\cite{PDF_uncertainty}.

Finally, uncertainties related to the limited number of simulated events are taken into account with the Barlow--Beeston-lite approach~\cite{Barlow-Beeston,Barlow-Beeston-Conway}. They are considered for all bins of the distributions that are used to extract the results. They are uncorrelated across the different samples and across the bins of a single distribution.

\begin{table}[tb!]
\centering
\topcaption{
The sources of uncertainty considered, categorized as to whether they affect the rate or shape of the distributions.
``s.d.'' refers to the standard deviation of the input variable and ``(mis)ID'' stands for ``(mis)identification''.
}
\label{tab:systematics}
\begin{tabular}{lrrcrr}
  \multirow{2}{*}{Source} & \multicolumn{5}{c}{Channel} \\
                                       \cline{2-6}
                                     & \etau & \mutau & \ditau & \emu & \mumu \\
\hline
  \textit{Rate}\\
  Integrated luminosity            & \multicolumn{5}{c}{1.2--2.5\%} \\
  Electron ID                      & 2\%     & \NA    & \NA    & 2\%  & \NA \\
  Electron trigger                 & 2\%     & \NA    & \NA    & \NA  & \NA \\
  Muon ID                          & \NA     & 2\%    & \NA    & 2\%  & 2\% \\
  Muon trigger                     & \NA     & 2\%    & \NA    & 2\%  & 2\% \\
  \Pe misID as \tauh               & 12\%    & \NA    & 12\%   & \NA  & \NA \\
  \PGm misID as \tauh              & \NA     & 25\%   & 25\%   & \NA  & \NA \\
  QCD multijet                     & \NA     & \NA    & \NA    & 20\% & 20\% \\
  \Wjets cross section             & \NA     & \NA    & \NA    & 6\%  & 6\% \\
  \DYjets cross section            & \multicolumn{5}{c}{20\% in \geqb, 3\% otherwise} \\
  \ttbar cross section             & \multicolumn{5}{c}{5.5\%} \\
  Diboson cross section            & \multicolumn{5}{c}{6\%} \\
  Single \PQt quark cross section  & \multicolumn{5}{c}{5.5\%} \\
  \Vgjets cross section            & \multicolumn{5}{c}{5\%} \\
  Electroweak {\PW}/{\PZ} boson cross section
                                   & \multicolumn{5}{c}{10\%} \\
  Triboson cross section           & \multicolumn{5}{c}{25\%} \\
  Jet energy scale                 & \multicolumn{5}{c}{5\% in 0j} \\
  \ptmiss scale                    & \multicolumn{5}{c}{Up to 4\%} \\[\cmsTabSkip]
  \textit{Shape}\\
  \tauh trigger                    & \NA     & \NA    & $\pm$1 s.d. in the SF            & \NA & \NA \\
  \tauh ID efficiency              & \multicolumn{3}{c}{$\pm$1 s.d. in SF, \pt extrapolation} & \NA & \NA \\
  \tauh energy scale               & \multicolumn{3}{c}{$\pm$1 s.d. on the energy scale} & \NA & \NA \\
  Energy scale \PGm misID as \tauh & \multicolumn{3}{c}{$\pm$1\% on the energy scale}    & \NA & \NA \\
  Energy scale \Pe misID as \tauh  & \multicolumn{3}{c}{$\pm$1 s.d. on the energy scale} & \NA & \NA \\
  FF shape variations              & \multicolumn{3}{c}{Syst. shape variations}          & \NA & \NA \\
  \PQb tagging efficiency          & \multicolumn{5}{c}{$\pm$1 s.d. in \PQb tagging SFs} \\
  \PQb tagging mistag rate         & \multicolumn{5}{c}{$\pm$1 s.d. in \PQb tagging SFs} \\
  Jet energy scale                 & \multicolumn{5}{c}{$\pm$1 s.d. in SF in 0b, \geqb} \\
  Jet energy resolution            & \multicolumn{5}{c}{$\pm$1 s.d. in SF in 0b, \geqb} \\
  ECAL trigger timing              & \multicolumn{5}{c}{$\pm$1 s.d. in SF} \\
  PDF variations                   & \multicolumn{5}{c}{Envelope of PDF variations} \\
  \muR \& \muF variations          & \multicolumn{5}{c}{Envelope of scale variations} \\
  \PZ boson \pt reweighting        & \multicolumn{5}{c}{Weight applied $\pm$50\%} \\
  \PQt quark \pt reweighting       & \multicolumn{5}{c}{Ref.~\cite{CMS:2021iuw} with larger variations} \\
\end{tabular}
\end{table}

All systematic uncertainties are summarized in Table~\ref{tab:systematics}. In addition to the statistical uncertainty due to the limited size of the simulated samples, the systematic uncertainties in the identification of genuine \tauh candidates and in the estimation of the \jtf contribution from data in CRs have the dominant impact on the expected sensitivity of this search.

\section{Results} \label{sec:results}

A binned maximum likelihood fit~\cite{CLs3} of the observed \STMET and $\chi$ distributions in all channels, event categories, and data-taking periods is used to search for a possible signal over the expected background. The nuisance parameters assigned to systematic uncertainties in the normalization of the distributions are modeled with log-normal probability distributions in the fit. Nuisance parameters for uncertainties affecting the shape of a distribution are modeled with Gaussian distributions.
In addition to searching for each \LQ production mechanism separately, results for the combined \LQ signal with a total cross section $\sigma^{\text{tot}}$ are also presented.

\begin{figure}[p!]
  \centering
  \includegraphics[width=\cmsFigWidth]{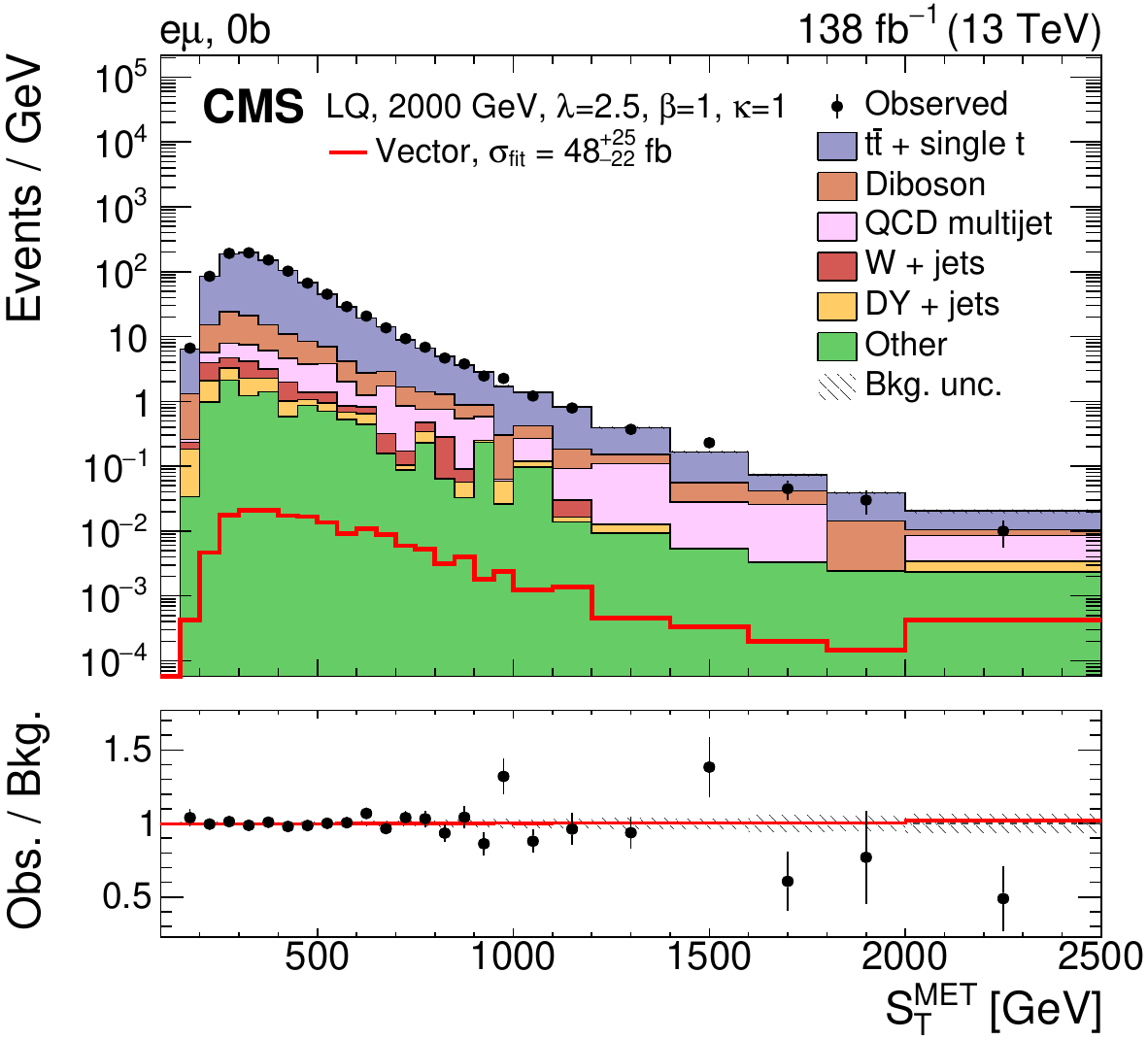}
  \includegraphics[width=\cmsFigWidth]{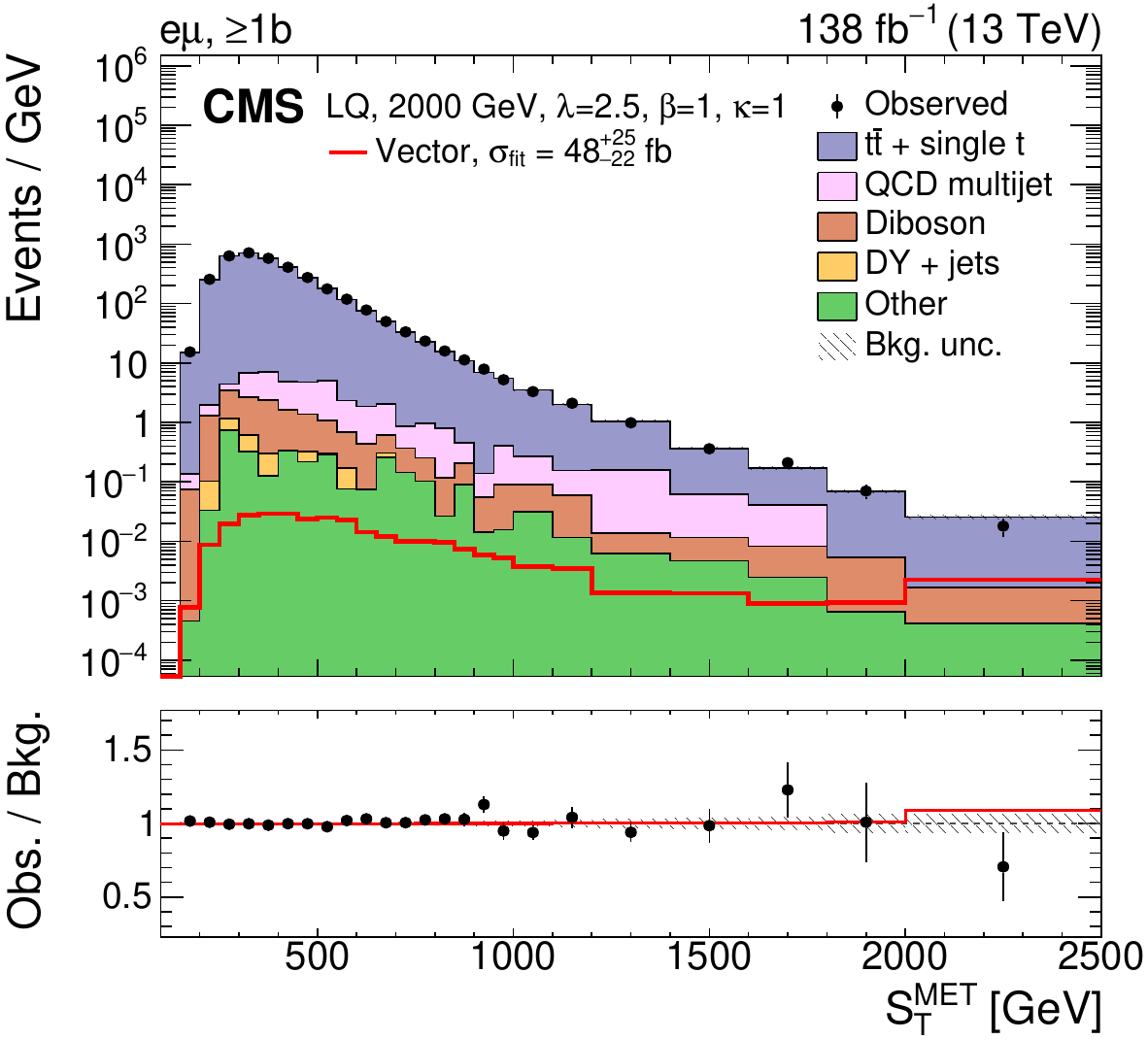}
  \includegraphics[width=\cmsFigWidth]{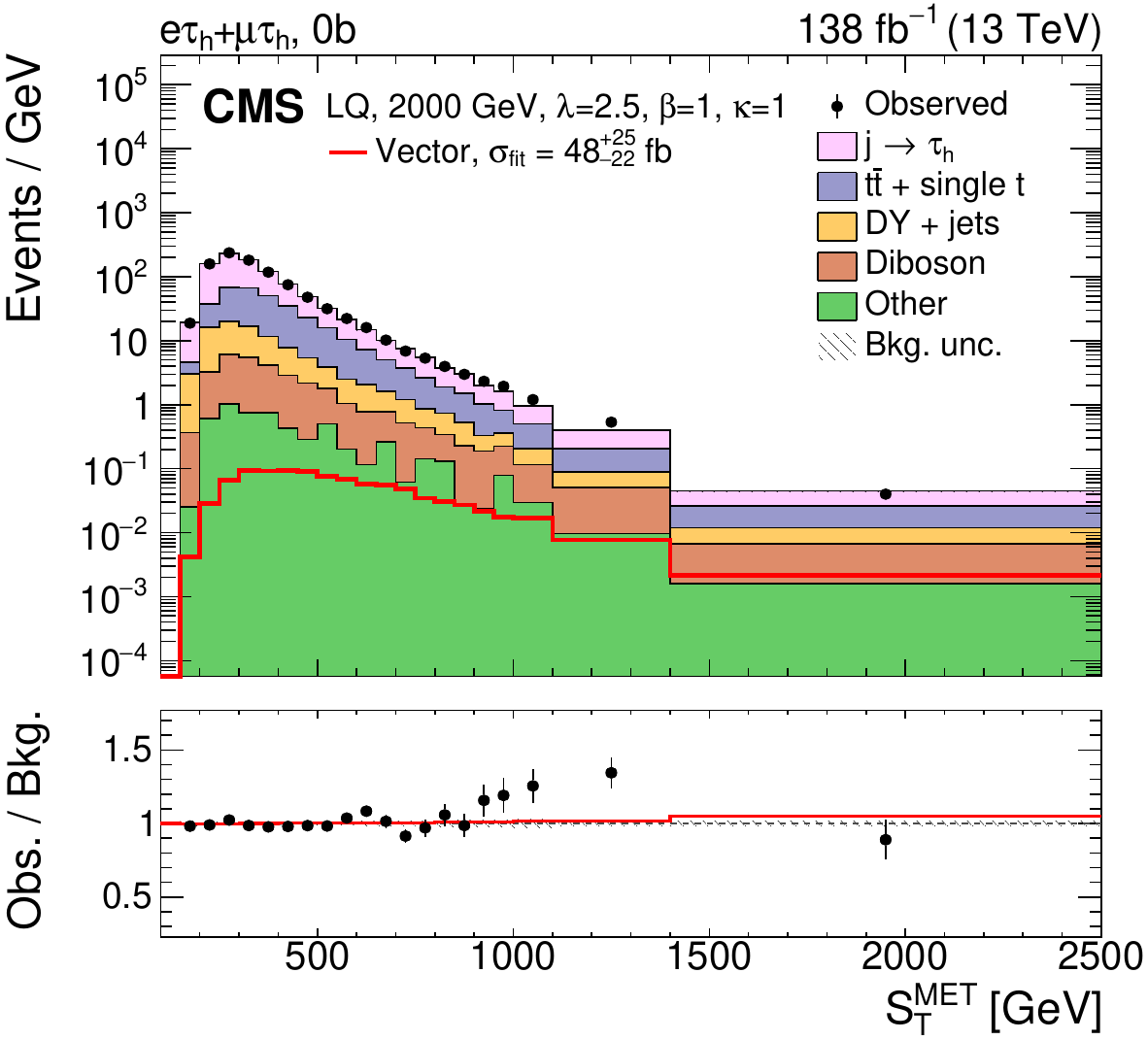}
  \includegraphics[width=\cmsFigWidth]{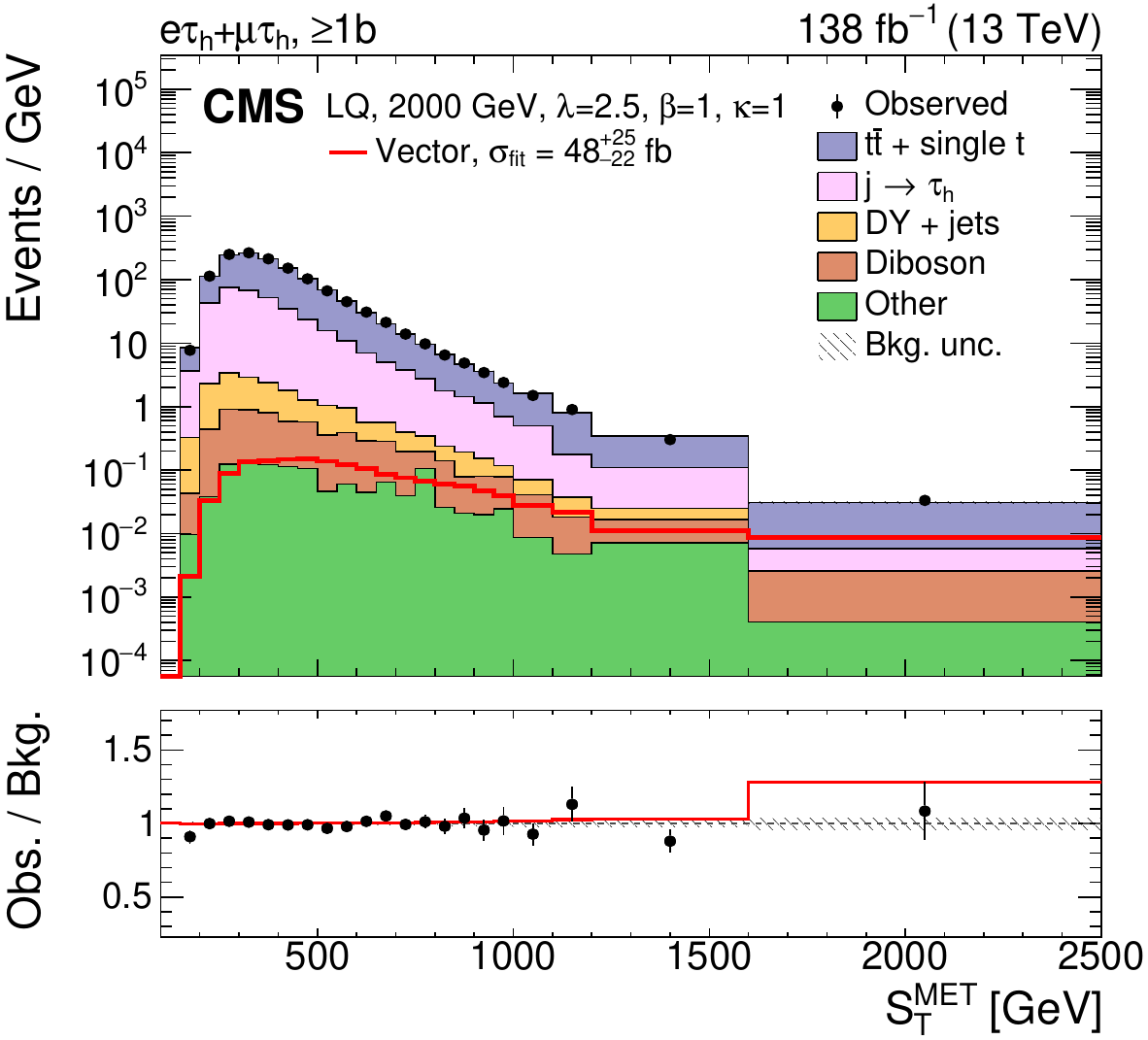}
  \includegraphics[width=\cmsFigWidth]{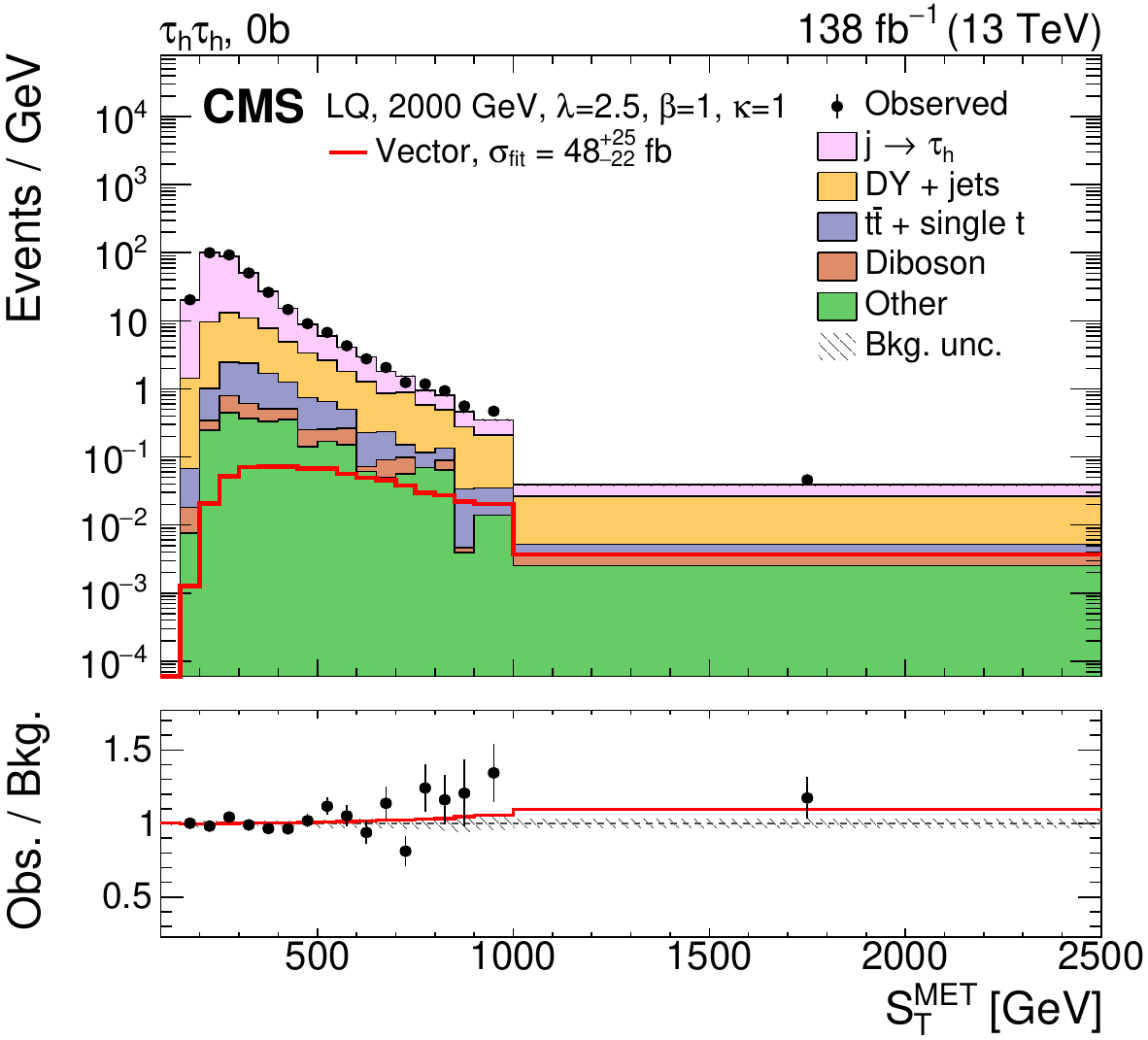}
  \includegraphics[width=\cmsFigWidth]{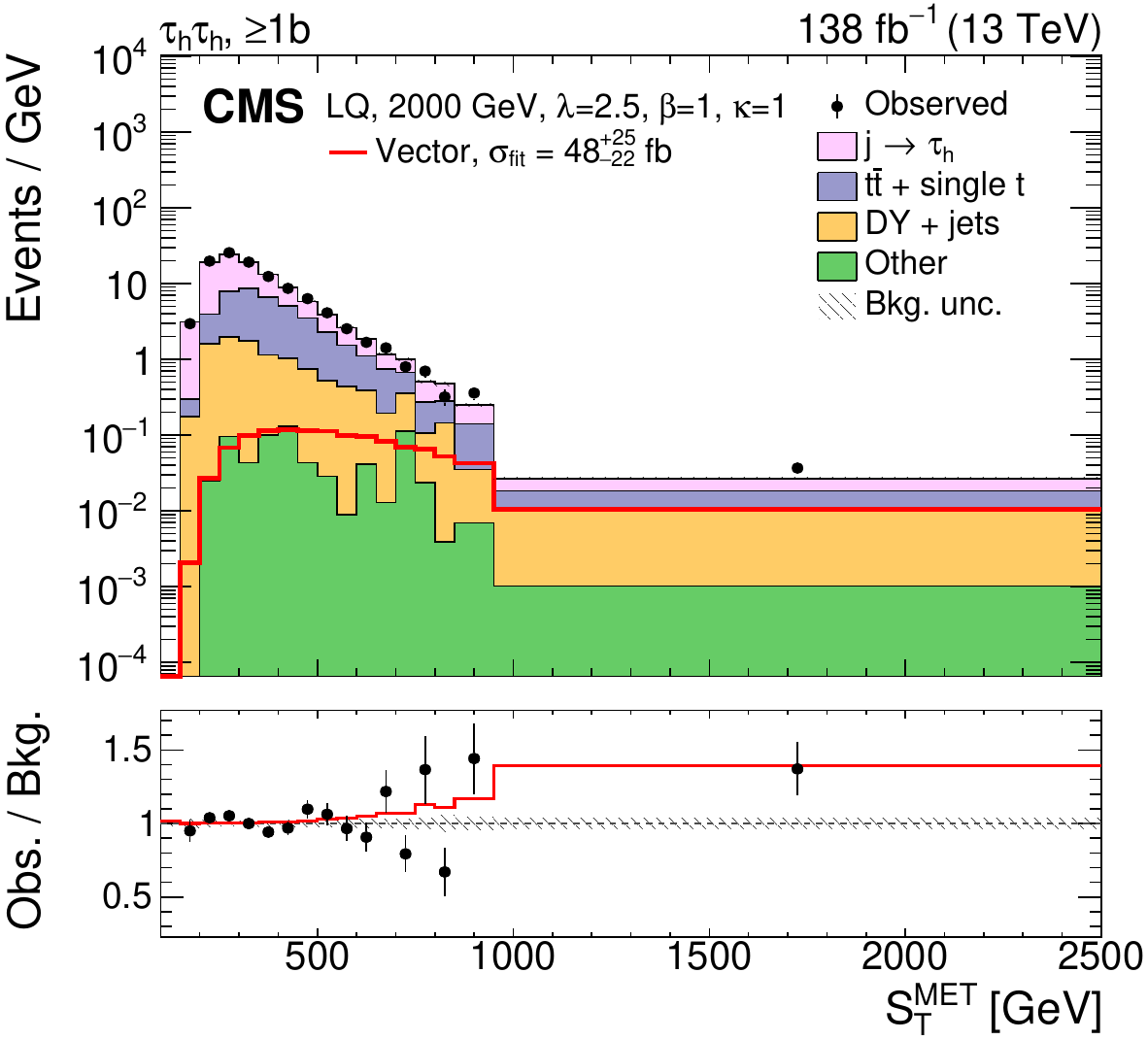}
  \caption{
Postfit distributions of \STMET for the combined 2016--2018 data set
after a simultaneous fit of the background and vector \LQ signal to the data.
The last bin includes the overflow. 
The \emu (top), \ltau (center), and \ditau (bottom) channels in the 0b (left) and \geqb (right) categories are shown.
The fitted signal distribution for the total vector \LQ signal (red line) with a mass of 2000\GeV, $\lambda = 2.5$, and $\kappa=1$ is overlaid.
In each distribution, the lower panel shows the ratio of the data (black markers) or the sum of the postfit signal and background (red line) to the postfit background.
The hatched band indicates the total postfit uncertainty in the background.
  }
  \label{fig:postfit_res}
\end{figure}

\begin{figure}[p!]
  \centering
  \includegraphics[width=\cmsFigWidth]{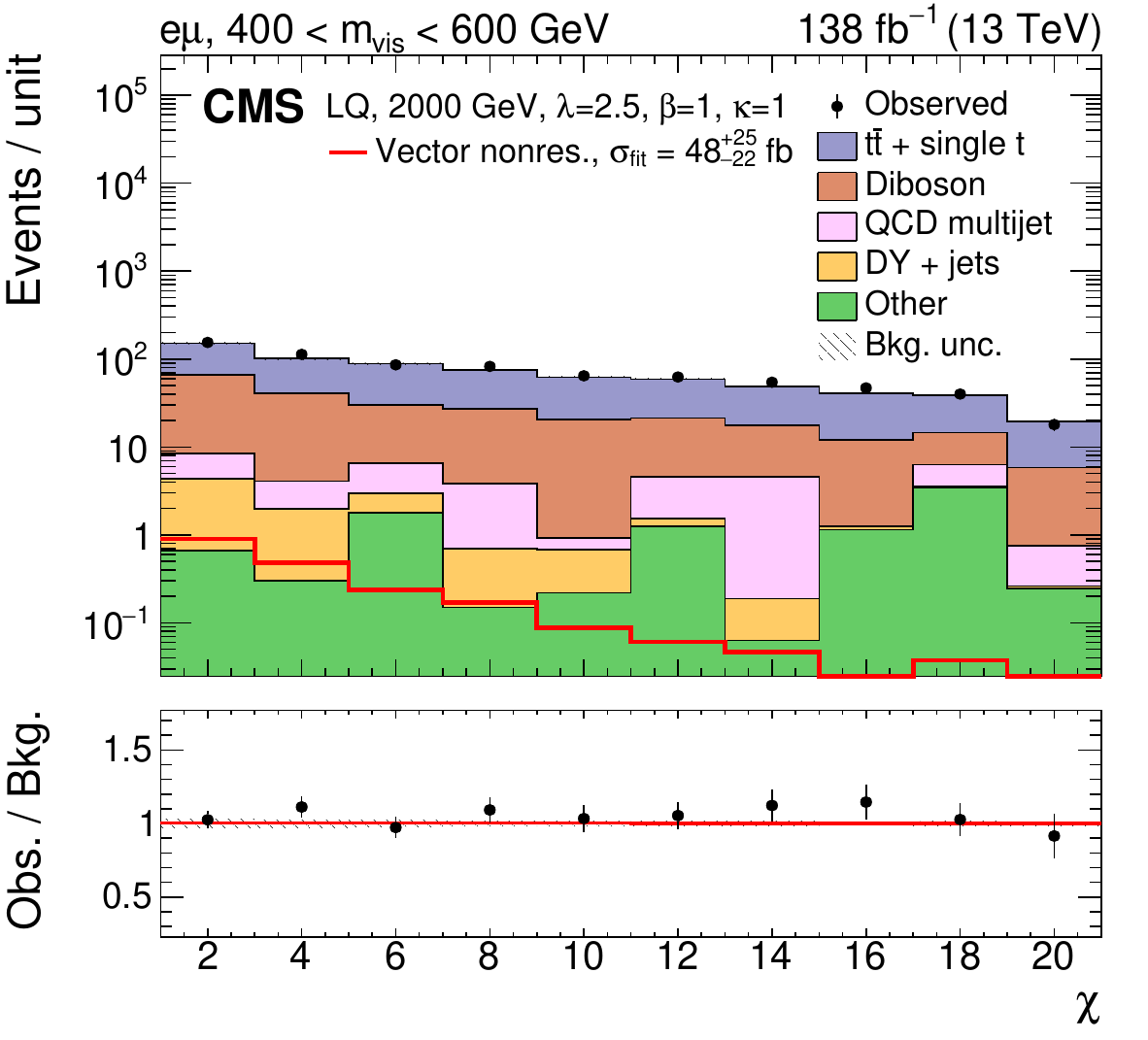}
  \includegraphics[width=\cmsFigWidth]{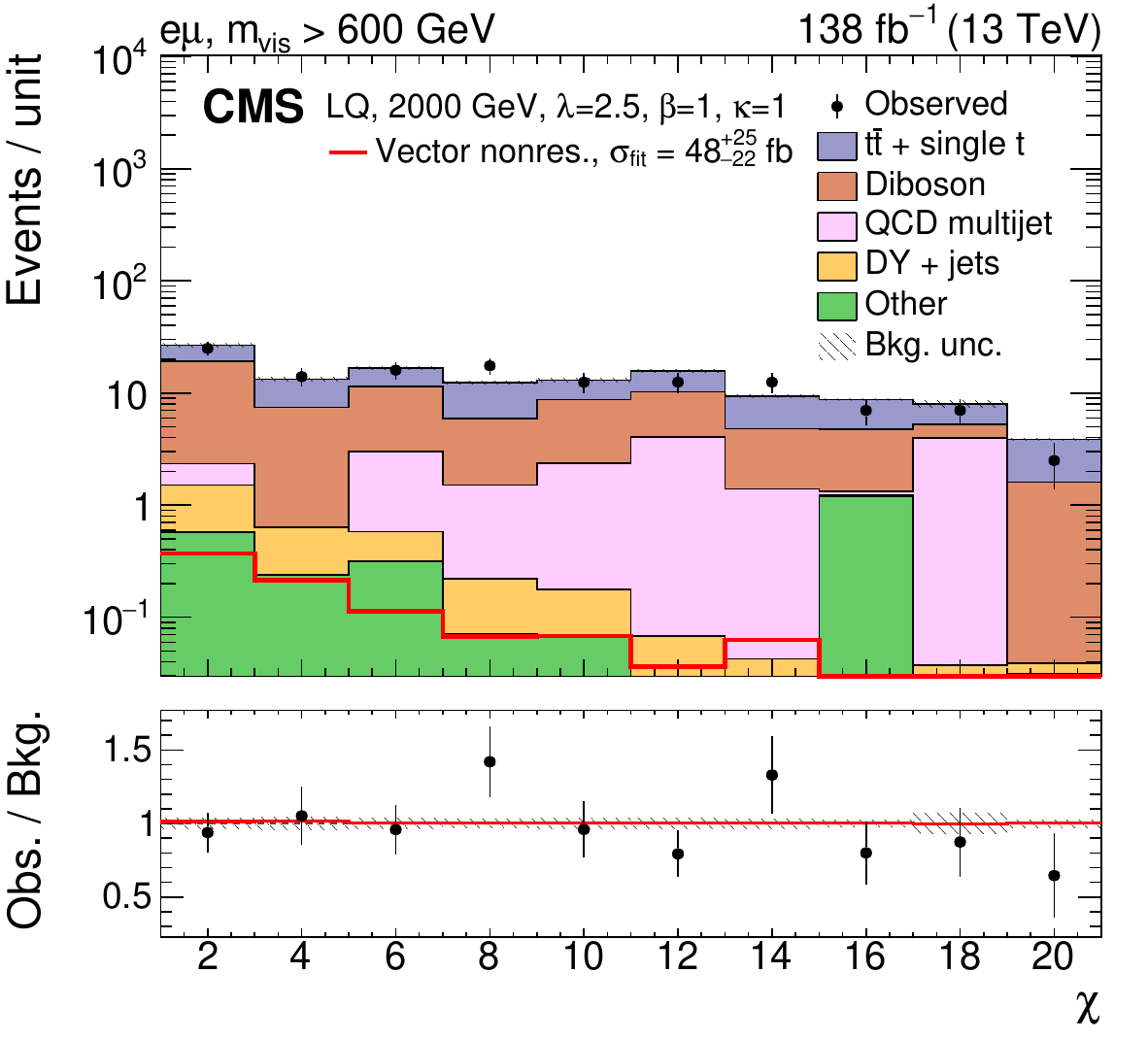}
  \includegraphics[width=\cmsFigWidth]{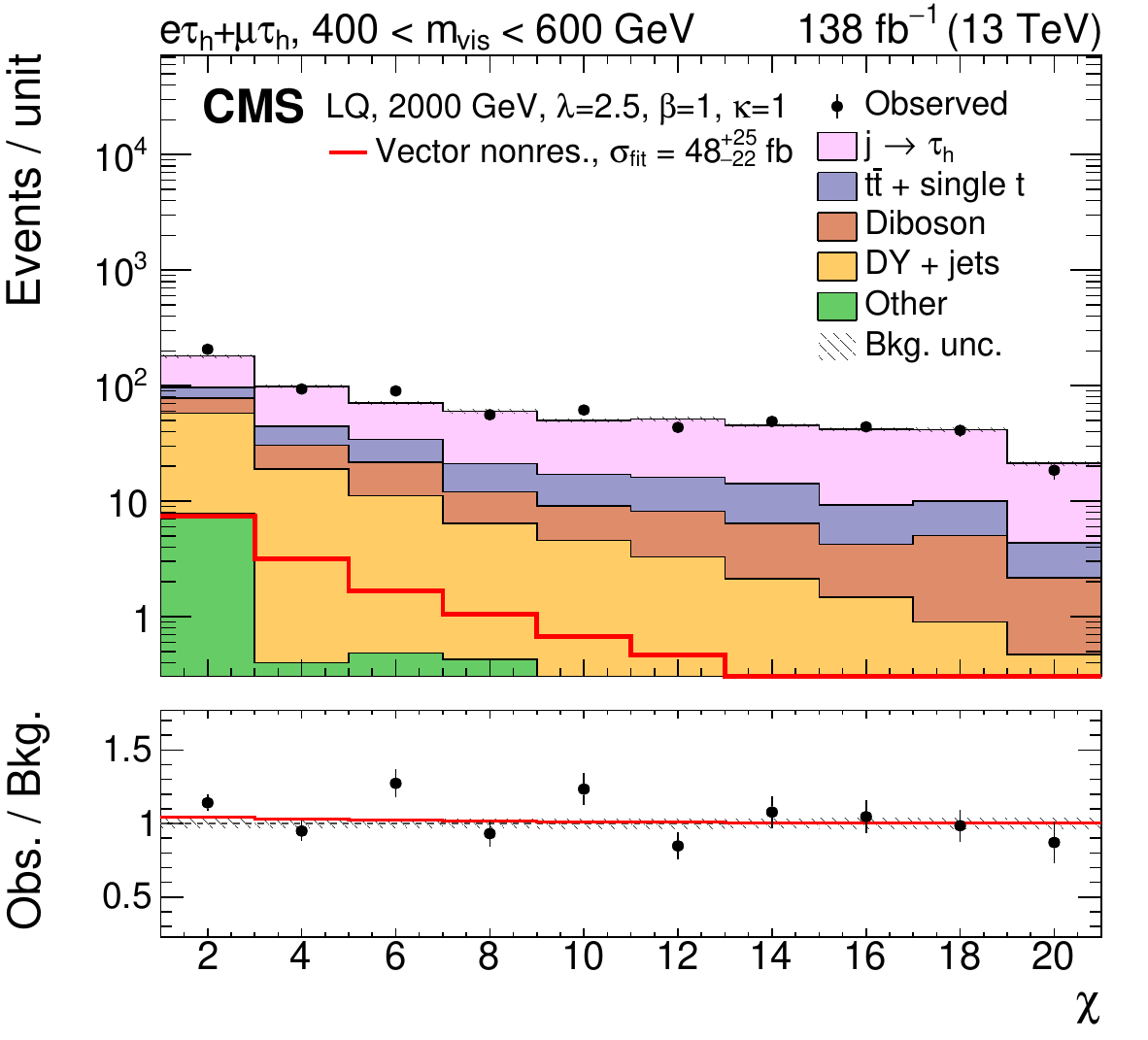}
  \includegraphics[width=\cmsFigWidth]{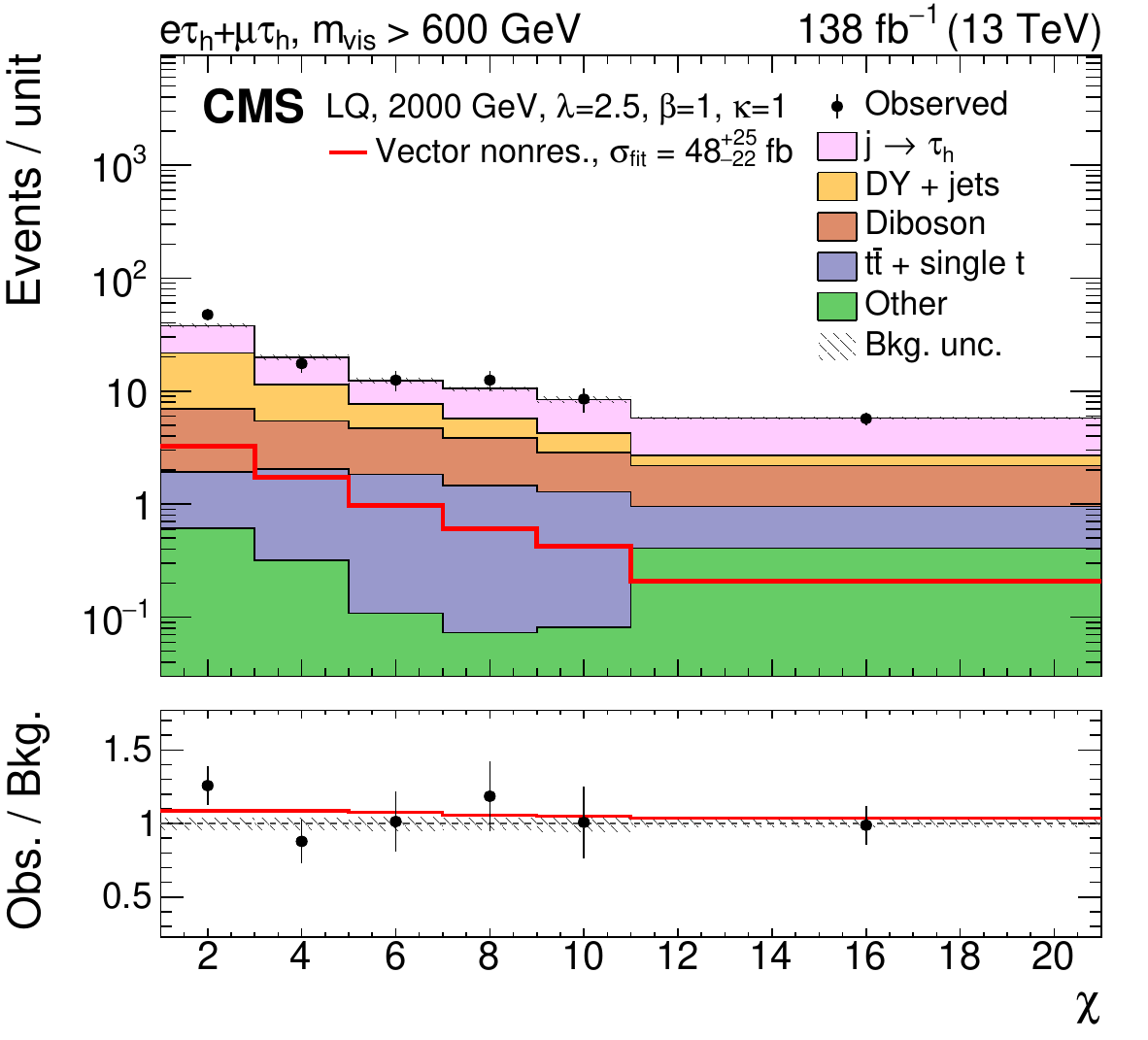}
  \includegraphics[width=\cmsFigWidth]{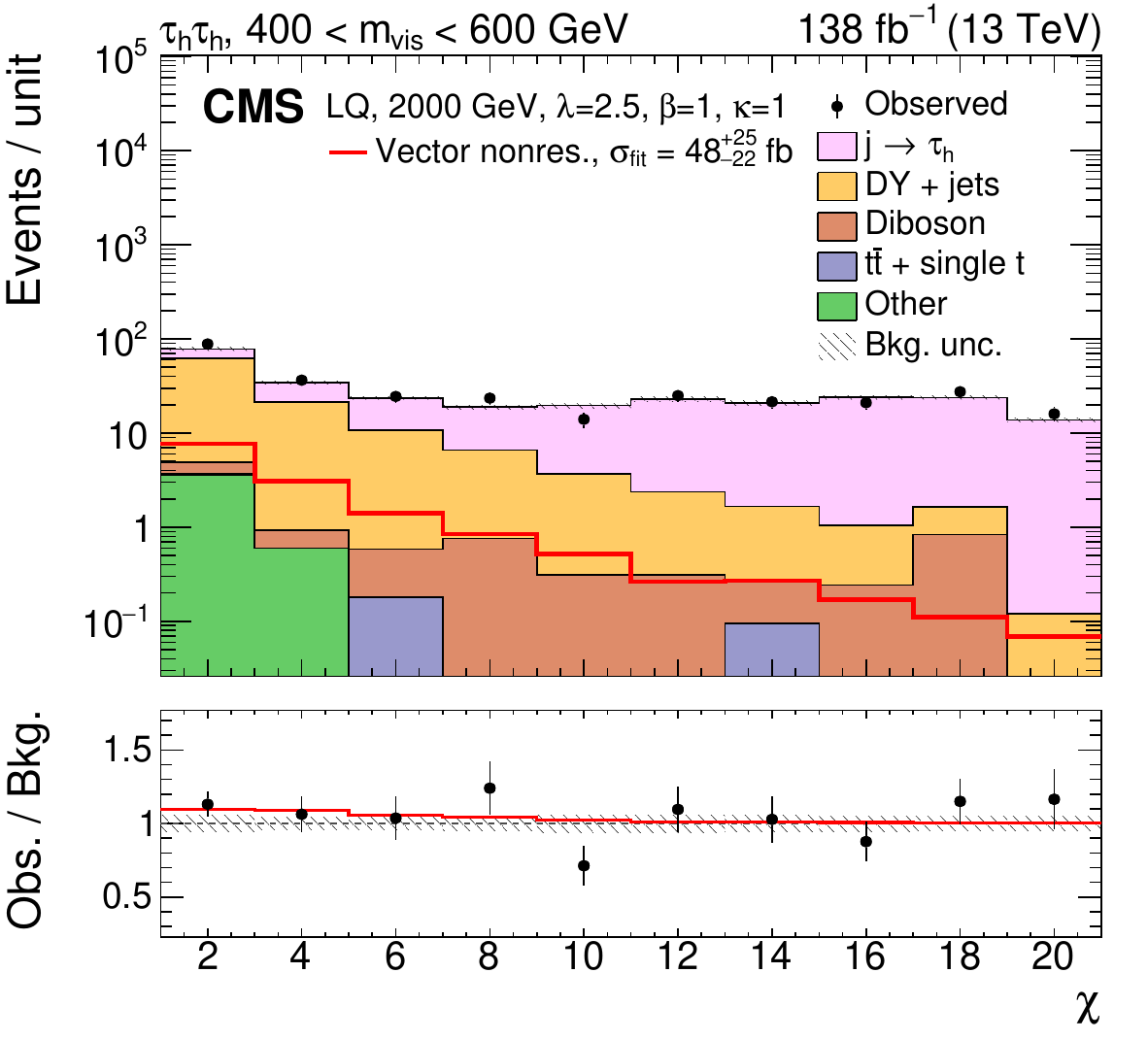}
  \includegraphics[width=\cmsFigWidth]{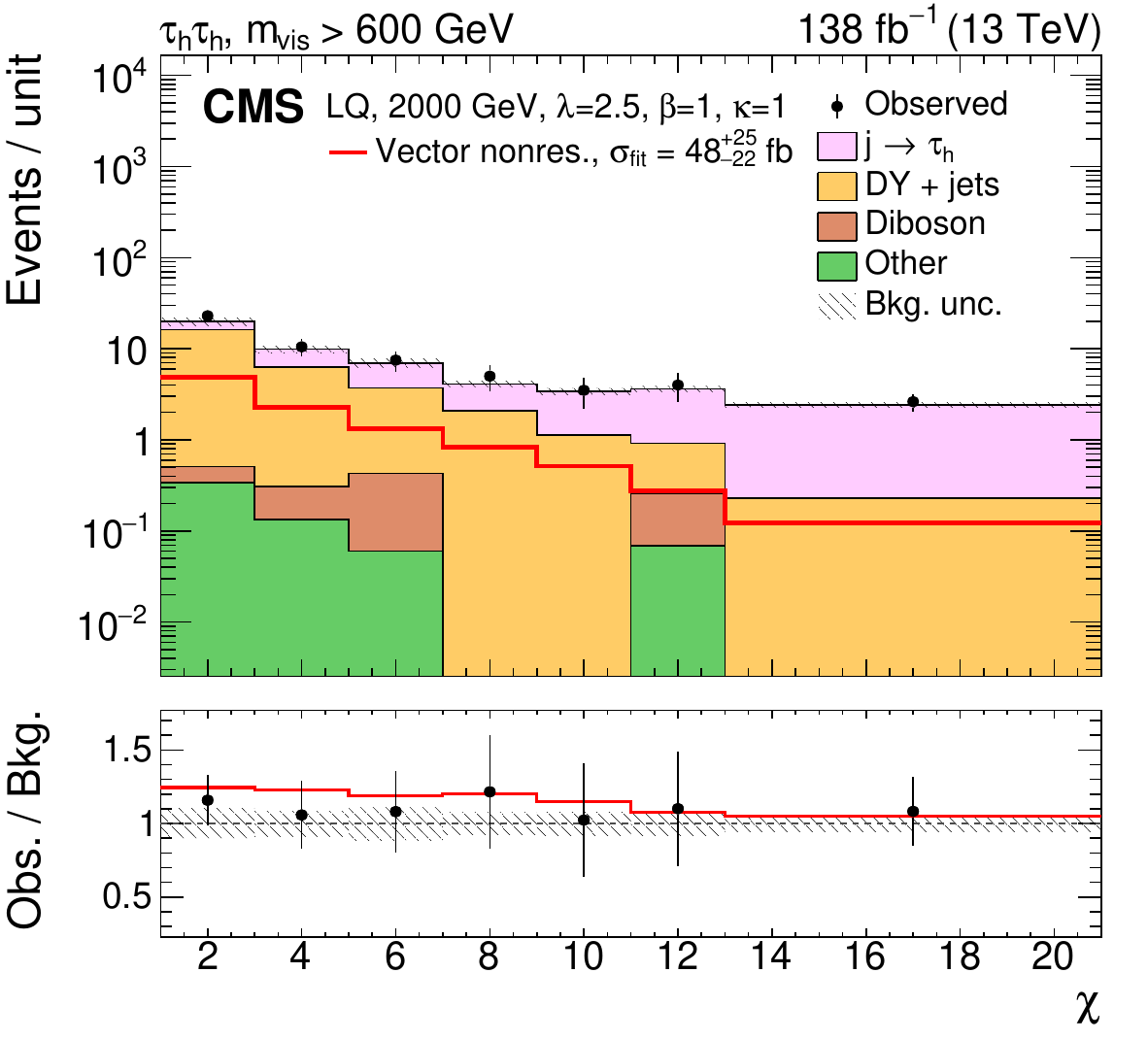}
  \caption{
Postfit distributions of $\chi$ for the combined 2016--2018 data set
after a simultaneous fit of the background and vector \LQ signal to the data.
The last bin includes the overflow.
The \emu (top), \ltau (center), and \ditau (bottom) channels in the $400<\mvis<600\GeV$ (left) and $\mvis>600\GeV$ (right) categories are shown.
The fitted signal distribution for the nonresonant vector \LQ model (red line) with a mass of 2000\GeV, $\lambda = 2.5$, and $\kappa=1$ is overlaid. The contribution from resonant \LQ production is neglected.
In each distribution, the lower panel shows the ratio of the data (black markers) or the sum of the postfit signal and background (red line) to the postfit background.
The hatched band indicates the total postfit uncertainty in the background.
  }
  \label{fig:postfit_0j}
\end{figure}

Figure~\ref{fig:postfit_res} shows representative postfit distributions of \STMET in the \ditau, \ltau, and \emu channels in the 0b and \geqb categories, while Fig.~\ref{fig:postfit_0j} shows the postfit distributions of $\chi$ in the $400<\mvis<600$ and $\mvis>600\GeV$ bins of the 0j category.
The total vector \LQ signal for the benchmark $\mLQ=2000\GeV$ with $\lambda=2.5$ and $\kappa=1$ is overlaid and normalized to the best-fit cross section.
The binning of all histograms is chosen such that the total statistical uncertainty in the SM background expectation of a given bin does not exceed 20\%.

The postfit event yields of the signal and backgrounds are compared to the observed data in Table~\ref{tab:yields} for \tautau final states with at least one \tauh, which are most sensitive to the \LQ signal, and all event categories considered in the fit.
Table~\ref{tab:xsecs} shows the best-fit signal cross section with the corresponding local significance for different \LQ production modes. 
For lower masses and coupling strengths $\lambda$, the observed data agree with the SM expectation within approximately one standard deviation, and within two standard deviations for $\lambda < 1.5$ and all \LQ masses considered. At higher masses and coupling strengths, however, an excess with a local significance of up to 2.8 standard deviations above the SM background expectation is found. 
The significance is computed using the ratio of the profile likelihood of the background-only fit to that of the fit in which the normalization of the total signal is a free parameter~\cite{Cowan:2010js}.
This excess is observed across several channels, as well as in both the $\chi$ and \STMET distributions in the different jet categories. The excess is most prominent for the nonresonant signal in the analysis, which contributes to the combined excess increasingly at larger \LQ masses and couplings. The significance of the nonresonant signal by itself has no strong dependence on \mLQ or $\lambda$, hence the look-elsewhere effect is negligible for the nonresonant signal. When combined with the pair and single production contributions to the total \LQ signal, the excess at lower \LQ masses and couplings becomes less significant.

\begin{sidewaystable}[p!]
  \centering
  \topcaption{
Event yields of data as well as signal and background after the combined fit to the data in \tautau final states with at least one \tauh.
The vector \LQ benchmark with $\mLQ=2000\GeV$, $\lambda=2.5$, and $\kappa=1$ is used as signal and normalized to the best-fit cross section of $\sigmafit=48_{-22}^{+25}\unit{fb}$.
The uncertainties reported are the total postfit uncertainties.
  }
  \label{tab:yields}
\begin{tabular}{llllllllllll}
  Category & Channel & Data & Signal & Total bkg. & \ttbar + single \PQt & \DYjets & Diboson & \jtf & Other \\ \hline
  \multirow{2}{*}{0b}
    & \ltau     & 47277  & $52 \pm 24$  & $47280 \pm 240$ & $12100 \pm 160$ & $3225 \pm 95 $ & $1298 \pm 71$ & $30380 \pm 300$ & $275 \pm 17$ \\
    & \ditau    & 16757  & $42 \pm 17$  & $16670 \pm 180$ & $389 \pm 14   $ & $2480 \pm 100$ & $75 \pm 4   $ & $13590 \pm 200$ & $134 \pm 8$ \\ [\cmsTabSkip]
  \multirow{2}{*}{\geqb}
    & \ltau     & 65810  & $92 \pm 44$  & $65780 \pm 280$ & $48160 \pm 380$ & $641 \pm 31  $ & $271 \pm 18 $ & $16650 \pm 370$ & $58 \pm 6$ \\
    & \ditau    & 5428   & $77 \pm 35$  & $5328 \pm 82  $ & $1644 \pm 57  $ & $512 \pm 26  $ & \NA           & $3139 \pm 98  $ & $33 \pm 4$ \\ [\cmsTabSkip]
  \multirow{2}{*}{$\mvis < 200\GeV$}
    & \ltau     & 24324  & $81 \pm 39$  & $24220 \pm 170$ & $3478 \pm 75  $ & $2068 \pm 42 $ & $1350 \pm 72$ & $17200 \pm 170$ & $114 \pm 7$ \\
    & \ditau    & 16357  & $67 \pm 29$  & $16290 \pm 150$ & $113 \pm 5    $ & $2112 \pm 88 $ & $52 \pm 3   $ & $13940 \pm 160$ & $72 \pm 5$\\ [\cmsTabSkip]
  \multirow{2}{*}{$\mvis\in [400,600]\GeV$}
    & \ltau     & 1407   & $31 \pm 14$  & $1331 \pm 33  $ & $182 \pm 9    $ & $195 \pm 6   $ & $142 \pm 8  $ & $793 \pm 33   $ & $20 \pm 4$ \\
    & \ditau    & 596    & $29 \pm 13$  & $561 \pm 23   $ & $1 \pm 0      $ & $205 \pm 13  $ & $9 \pm 1    $ & $337 \pm 19   $ & $8 \pm 3$ \\ [\cmsTabSkip]
  \multirow{2}{*}{$\mvis > 600\GeV$}
    & \ltau     & 254    & $16 \pm 7 $  & $234 \pm 11   $ & $20 \pm 2     $ & $59 \pm 3    $ & $43 \pm 3   $ & $106 \pm 8    $ & $6 \pm 2$ \\
    & \ditau    & 128    & $21 \pm 9 $  & $114 \pm 10   $ & \NA             & $60 \pm 6    $ & $2 \pm 0    $ & $52 \pm 8     $ & $1 \pm 0$ \\ [\cmsTabSkip]
  \end{tabular}
\end{sidewaystable}

\begin{table}[phtb!]
  \centering
  \topcaption{
Best-fit \LQ cross sections \sigmafit for various masses and coupling strengths $\lambda$,
and the corresponding local significance $z$ (given in standard deviations)
for different production modes individually, as well as their combination.
The look-elsewhere effect is negligible.
  }
  \label{tab:xsecs}
\renewcommand{\arraystretch}{1.3}
\begin{tabular}{@{\extracolsep{4pt}}llclc@{}}
  \multirow{2}{*}{Signal}
    & \multicolumn{2}{c}{$\mLQ=1400\GeV$} & \multicolumn{2}{c}{$\mLQ=2000\GeV$} \\[0.4mm]
  \cline{2-3} \cline{4-5} \rule{0pt}{10pt}
    & \sigmafit [fb] & $z$ & $\sigmafit$ [fb] & $z$ \\
\hline
  \textit{Scalar}\\
  Pair                    & $0.46_{-0.49}^{+0.51}$ & 1.0 & $0.39_{-0.43}^{+0.45}$ & 0.9 \\
  Single, $\lambda=1$     & $1.32_{-0.99}^{+1.03}$ & 1.3 & $0.84_{-0.70}^{+0.73}$ & 1.2 \\
  Single, $\lambda=2.5$   & $8.7_{-5.7}^{+6.0}   $ & 1.5 & $17_{-11}^{+12}      $ & 1.5 \\
  Nonres.                 & $92_{-38}^{+44}      $ & 2.8 & $83_{-34}^{+39}      $ & 2.8 \\
  Total, $\lambda=1$      & $2.3_{-2.0}^{+2.1}   $ & 1.2 & $10.3_{-6.4}^{+7.1}  $ & 1.6 \\
  Total, $\lambda=2.5$    & $47_{-22}^{+25}      $ & 2.4 & $78_{-32}^{+37}      $ & 2.8 \\[\cmsTabSkip]
  \textit{Vector, $\kappa=0$}\\
  Pair                    & $0.46_{-0.48}^{+0.51}$ & 1.0 & $0.41_{-0.42}^{+0.44}$ & 1.0 \\
  Single, $\lambda=1$     & $1.20_{-0.92}^{+0.96}$ & 1.3 & $0.81_{-0.68}^{+0.71}$ & 1.2 \\
  Single, $\lambda=2.5$   & $19_{-12}^{+13}      $ & 1.6 & $31_{-22}^{+22}      $ & 1.5 \\
  Nonres.                 & $71_{-29}^{+34}      $ & 2.8 & $62_{-26}^{+30}      $ & 2.7 \\
  Total, $\lambda=1$      & $1.8_{-1.6}^{+1.7}   $ & 1.1 & $8.2_{-5.2}^{+5.7}   $ & 1.6 \\
  Total, $\lambda=2.5$    & $47_{-21}^{+24}      $ & 2.5 & $62_{-26}^{+31}      $ & 2.7 \\[\cmsTabSkip]
  \textit{Vector, $\kappa=1$}\\
  Pair                    & $0.46_{-0.48}^{+0.51}$ & 1.0 & $0.41_{-0.42}^{+0.44}$ & 1.0 \\
  Single, $\lambda=1$     & $1.20_{-0.92}^{+0.96}$ & 1.3 & $0.81_{-0.68}^{+0.71}$ & 1.2 \\
  Single, $\lambda=2.5$   & $9.8_{-6.7}^{+7.0}   $ & 1.5 & $24_{-18}^{+19}      $ & 1.3 \\
  Nonres.                 & $71_{-29}^{+34}      $ & 2.8 & $62_{-26}^{+30}      $ & 2.7 \\
  Total, $\lambda=1$      & $0.72_{-0.72}^{+0.75}$ & 1.0 & $1.8_{-1.5}^{+1.6}   $ & 1.2 \\
  Total, $\lambda=2.5$    & $12.5_{-7.5}^{+8.3}  $ & 1.7 & $48_{-22}^{+25}      $ & 2.5 \\
\end{tabular}
\end{table}

To better understand the origin of the excess and how compatible the signal model is with each different final state, the fit of the total \LQ signal is performed again, but with one independent signal normalization parameter in each of the three jet categories (0j, 0b, \geqb) while keeping common nuisance parameters.
For each fitted cross section, a significance is computed while leaving the other two signal normalization parameters freely floating. 
For the $\mLQ=2000\GeV$, $\lambda=2.5$ benchmark, which yields the largest combined excess, the local significances in the 0j and \geqb categories range from 2.5 to 3.2 and 1.8 to 2.0 standard deviations, respectively, depending on the spin of the \LQ and the coupling parameter $\kappa$. In the 0b category, a local significance between 3.4 and 3.7 standard deviations is found for the scalar and vector \LQ models.

In a search performed by the ATLAS Collaboration~\cite{ATLAS:2023vxj}, the \LQ benchmark used in this analysis has been excluded, warranting the study of alternative signal models as a potential explanation of the results.
As can be seen in Fig.~\ref{fig:postfit_res}, the largest excess of events is observed in the 0b category at high values of \STMET and most prominently in the \ltau channel, which is not expected for the model with exclusive \LQ-\PQb-\PGt couplings considered in this analysis. An improved description of the data could be possible with additional \LQ couplings to particles of lighter generations.
An independent analysis group based on Ref.~\cite{HIG-21-001} finds consistent results when applying similar event categorization and selection. A detailed study of events in the 0b category with $\STMET > 800\GeV$ excluded detector-level and reconstruction issues as a cause of the excess. This study also led to an improved modeling of the background due to jets misidentified as \tauh objects with high \pt in this analysis, and the systematic uncertainties associated with genuine and misidentified \tauh candidates. However, the excess was found to be robust against alternative \tauh uncertainty models such as modified correlation schemes or different parametrizations of the uncertainties.

\begin{figure}[htb!]
  \centering 
  \includegraphics[width=\cmsFigWidth]{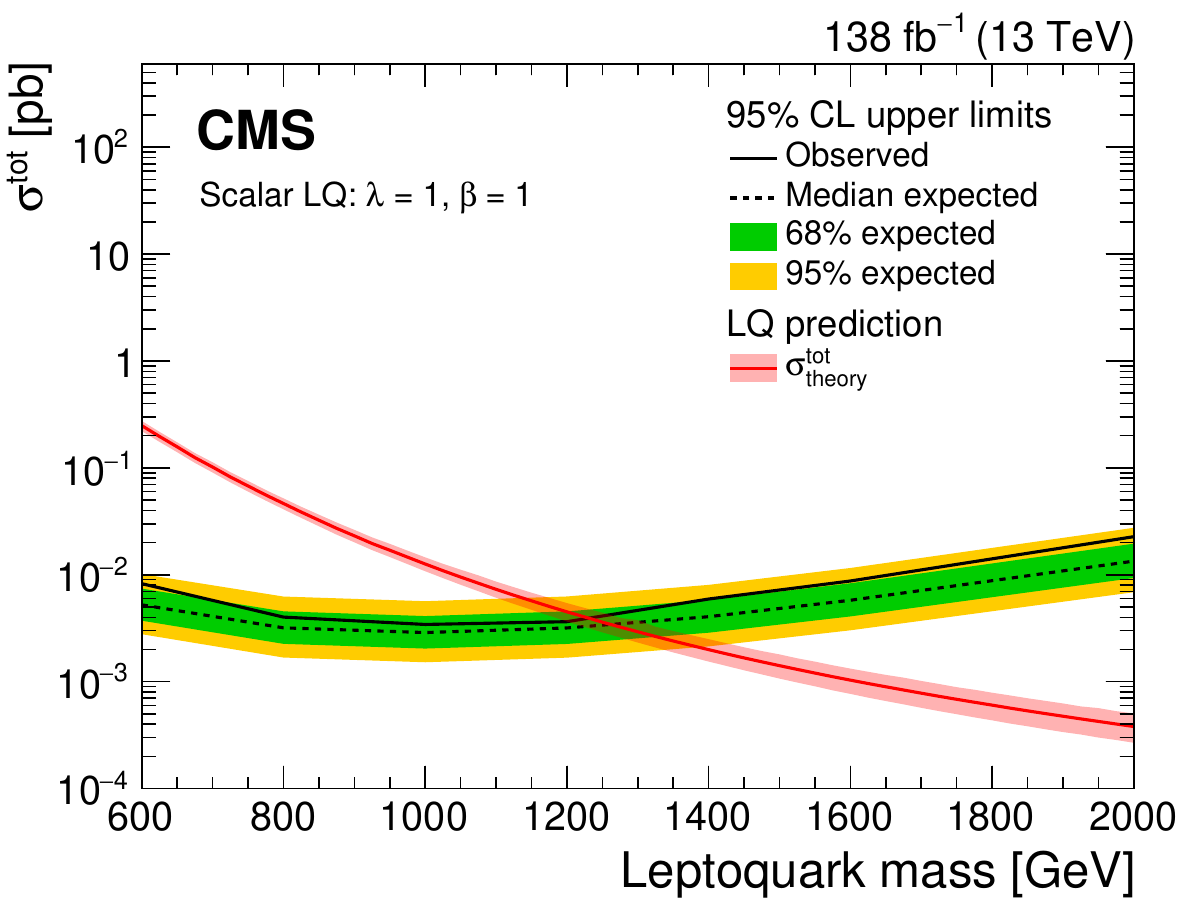}
  \includegraphics[width=\cmsFigWidth]{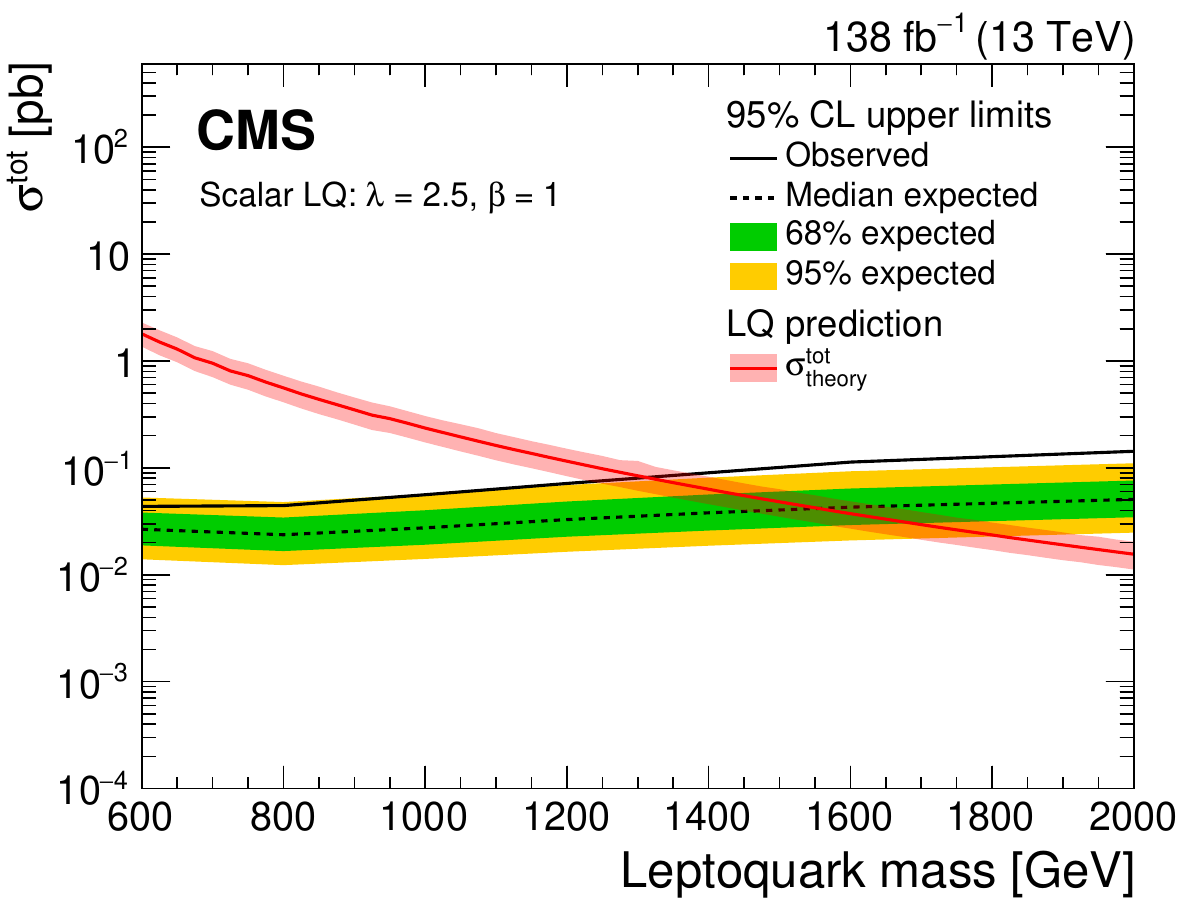}
  \caption{
Observed and expected upper limit on the total cross section of a scalar \LQ signal with $\lambda=1$ (left) and 2.5 (right) at the 95\% \CL under the assumption of exclusive \LQ couplings to \PQb quarks and \PGt leptons.
The inner (green) band and the outer (yellow) band indicate the regions containing 68 and 95\%,
respectively, of the distribution of limits expected under the background-only hypothesis.
The red line shows the cross section with the shaded band indicating the theoretical uncertainties.
  }
  \label{fig:limit_SLQ}
\end{figure}

\begin{figure}[p!]
  \centering 
  \includegraphics[width=\cmsFigWidth]{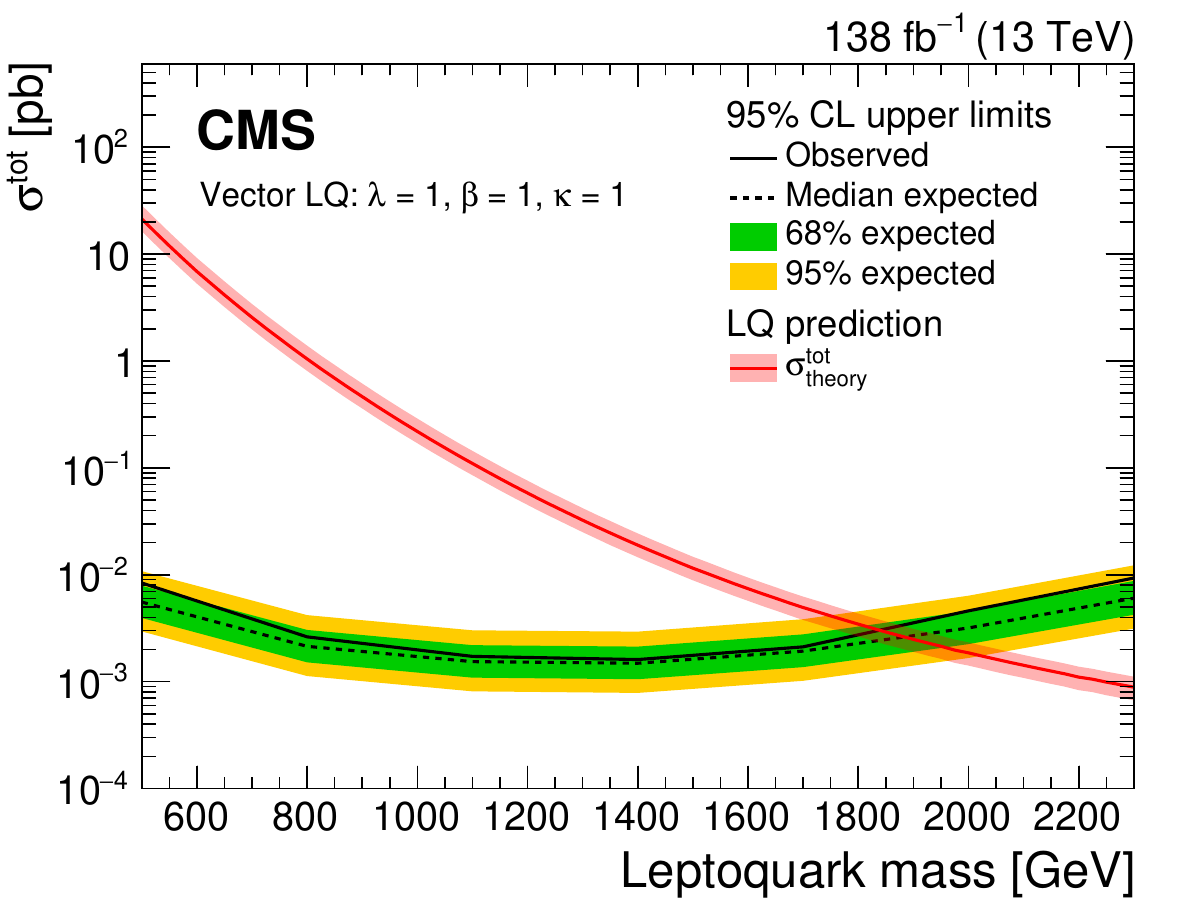}\quad
  \includegraphics[width=\cmsFigWidth]{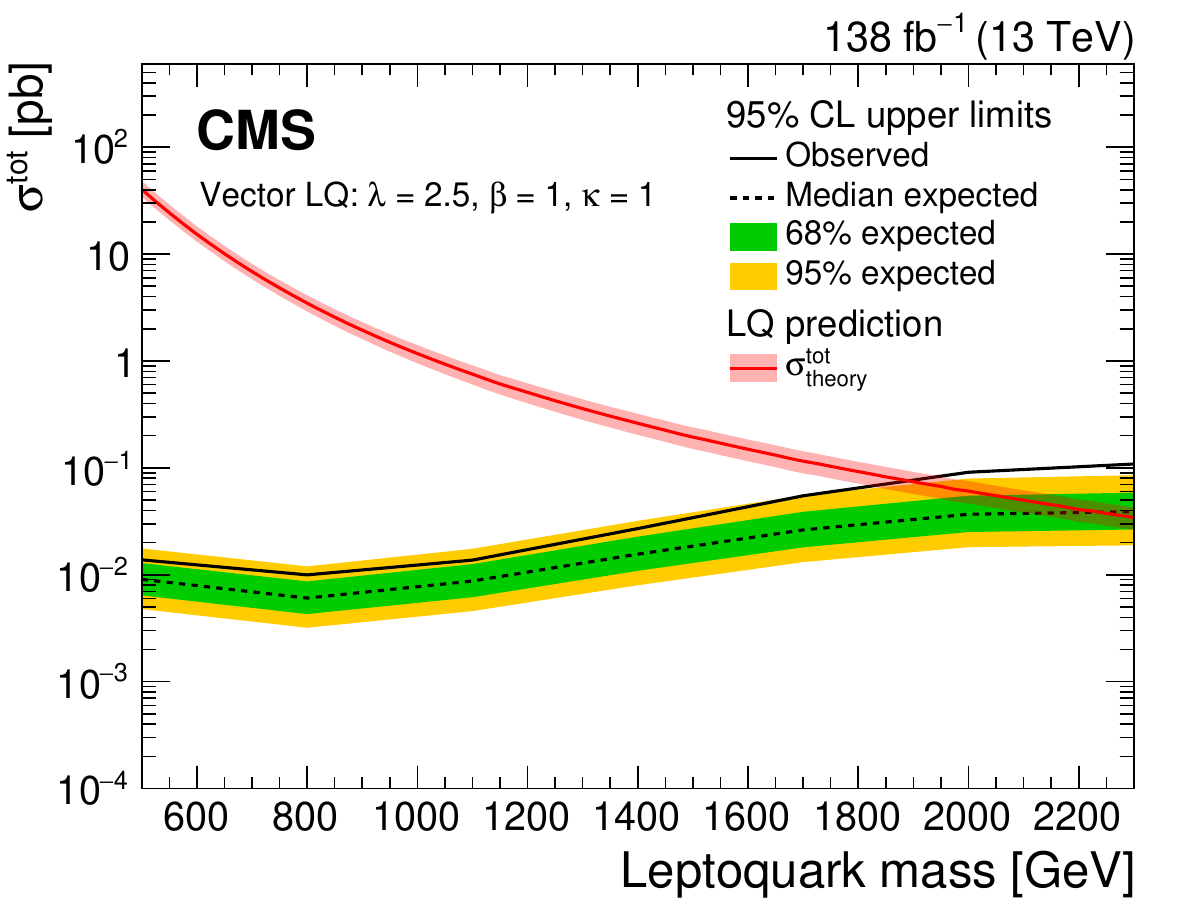}
  \includegraphics[width=\cmsFigWidth]{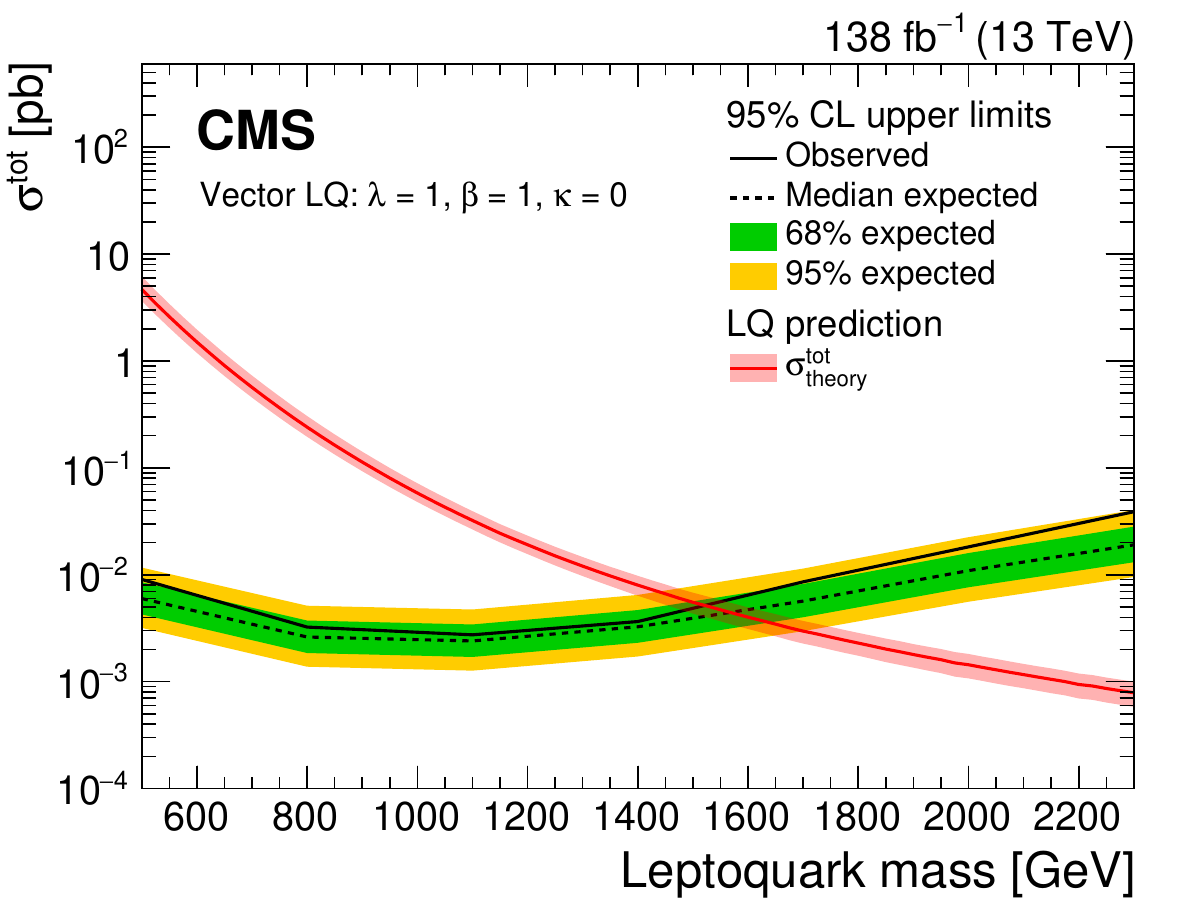}\quad
  \includegraphics[width=\cmsFigWidth]{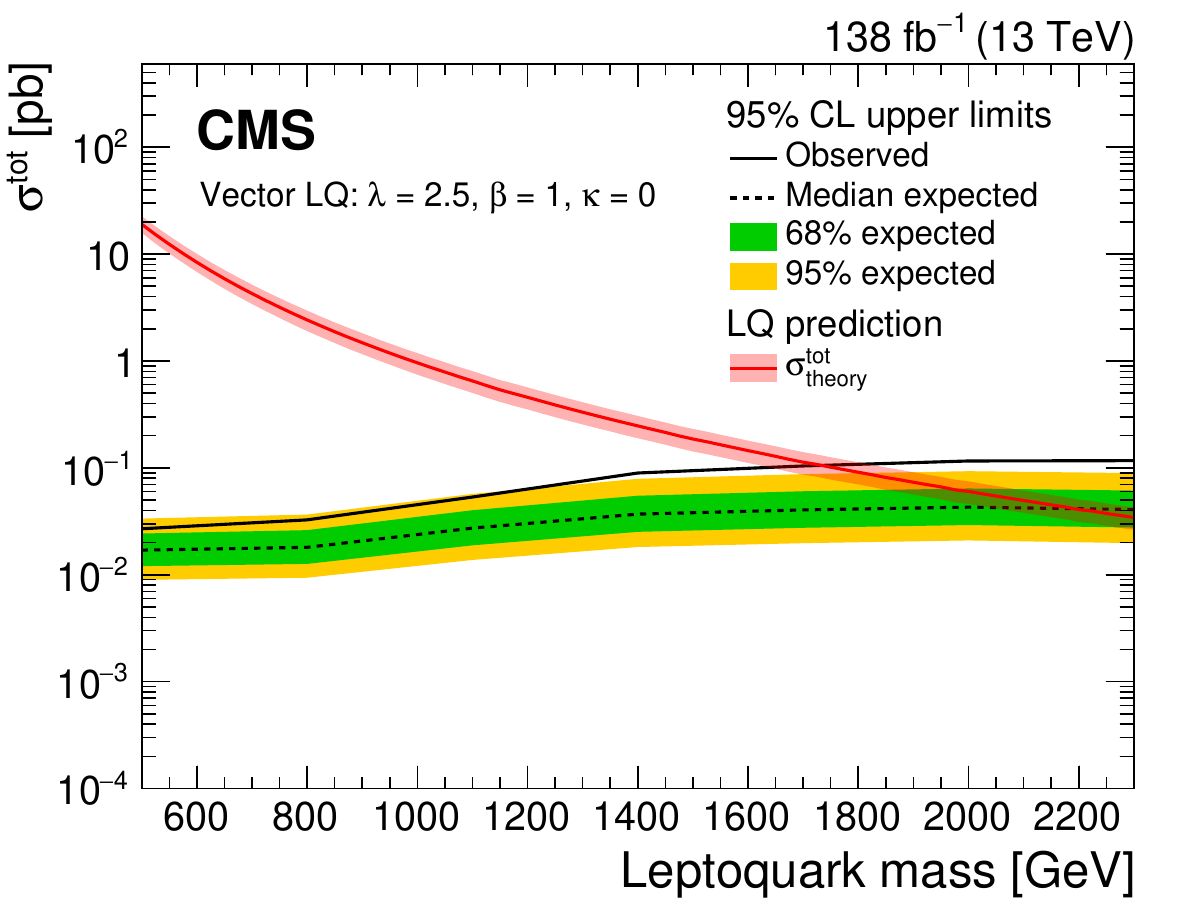}
  \caption{
Observed and expected upper limit on the total cross section of a vector \LQ signal with $\lambda=1$ (left) and 2.5 (right) at the 95\% \CL under the assumption of exclusive \LQ couplings to \PQb quarks and \PGt leptons.
The upper (lower) row assumes a coupling parameter $\kappa=1$\,(0).
The inner (green) band and the outer (yellow) band indicate the regions containing 68 and 95\%,
respectively, of the distribution of limits expected under the background-only hypothesis.
The red line shows the cross section calculated at LO with the shaded band indicating the theoretical uncertainties.
  }
  \label{fig:limit_VLQ}
\end{figure}

We set upper limits on the cross section of single, pair, and nonresonant production, as well as their sum, using a profile likelihood ratio test statistic and the \CLs criterion~\cite{CL, CLs1}. The distributions of the test statistic are determined analytically using asymptotic approximations~\cite{Cowan:2010js}.
All limits are placed under the assumption of \LQs coupling exclusively to \PQb quarks and \PGt leptons ($\beta = 1$).
The single production cross section approximately scales with $\lambda^2$ at lower values of $\lambda<1.5$, while the nonresonant production cross section is proportional to $\lambda^4$. Consequently, an upper limit on the total cross section $\sigma^\text{tot}$ can be derived for different values of $\lambda$. To exclude regions of the $\lambda$-\mLQ parameter space, the upper limit on $\lambda$ at the 95\% \CL is computed by varying $\lambda$. The observed and expected upper limits at the 95\% \CL on the total cross section of scalar and vector \LQ production are shown in Figs.~\ref{fig:limit_SLQ} and \ref{fig:limit_VLQ}, respectively, for $\lambda=1$ and 2.5.

\begin{figure}[htb!]
  \centering
  \includegraphics[width=1.1\cmsFigWidth]{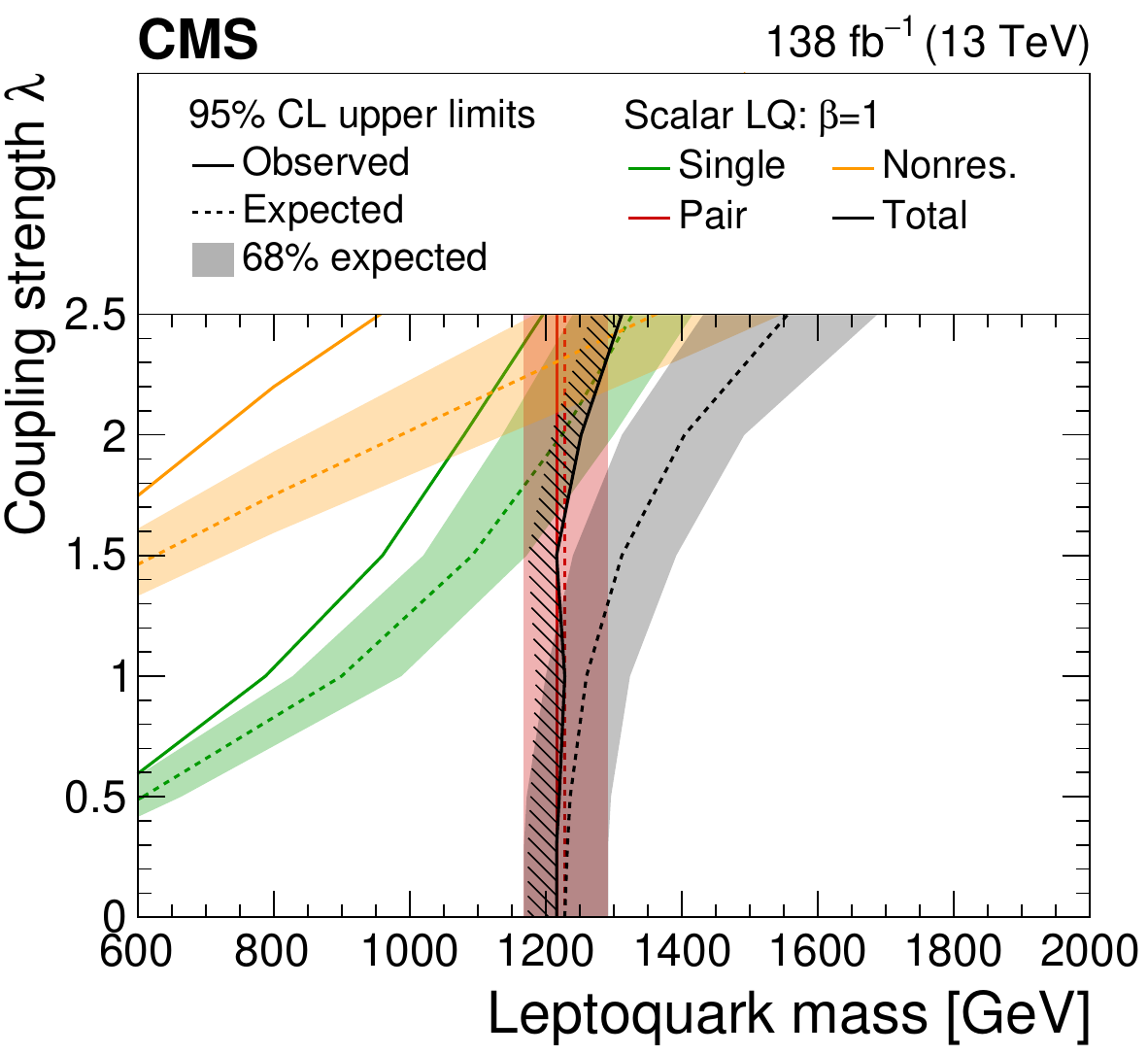}
  \caption{
Observed and expected upper limit at the 95\% \CL on the coupling strength $\lambda$ of a scalar \LQ under the assumption of exclusive \LQ couplings to \PQb quarks and \PGt leptons.
The limits derived for the single (green), pair (red), nonresonant (orange), and total \LQ production (black) are shown.
The shaded bands around the expected limit lines correspond to the regions containing 68\% of the distribution of limits expected under the background-only hypothesis. The hatches indicate the excluded side of the parameter space with respect to the combined observed limit.
   }
   \label{fig:limit_combined_SLQ_2D}
\end{figure}

\begin{figure}[htb!]
  \centering
  \includegraphics[width=1.1\cmsFigWidth]{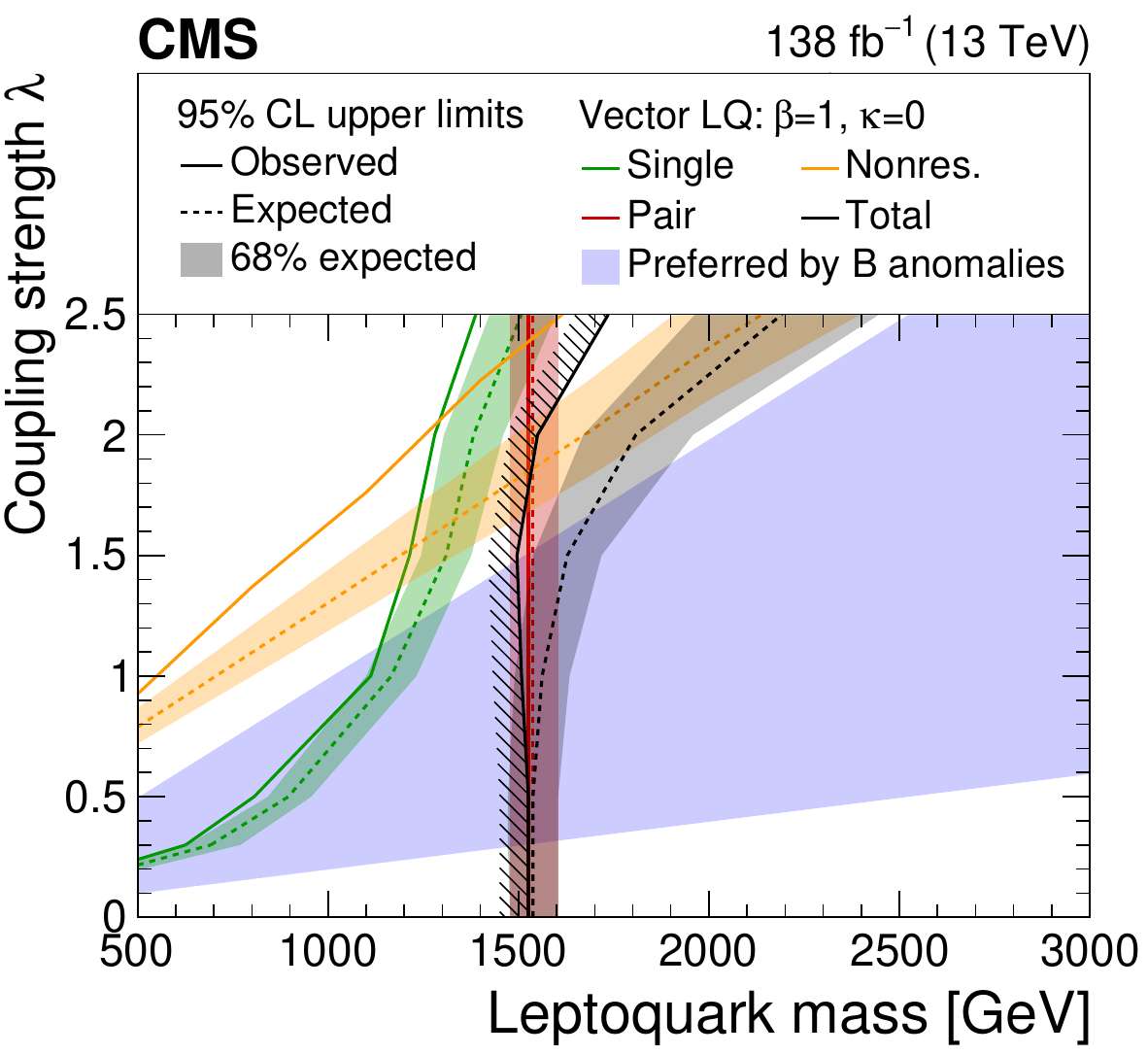}\quad
  \includegraphics[width=1.1\cmsFigWidth]{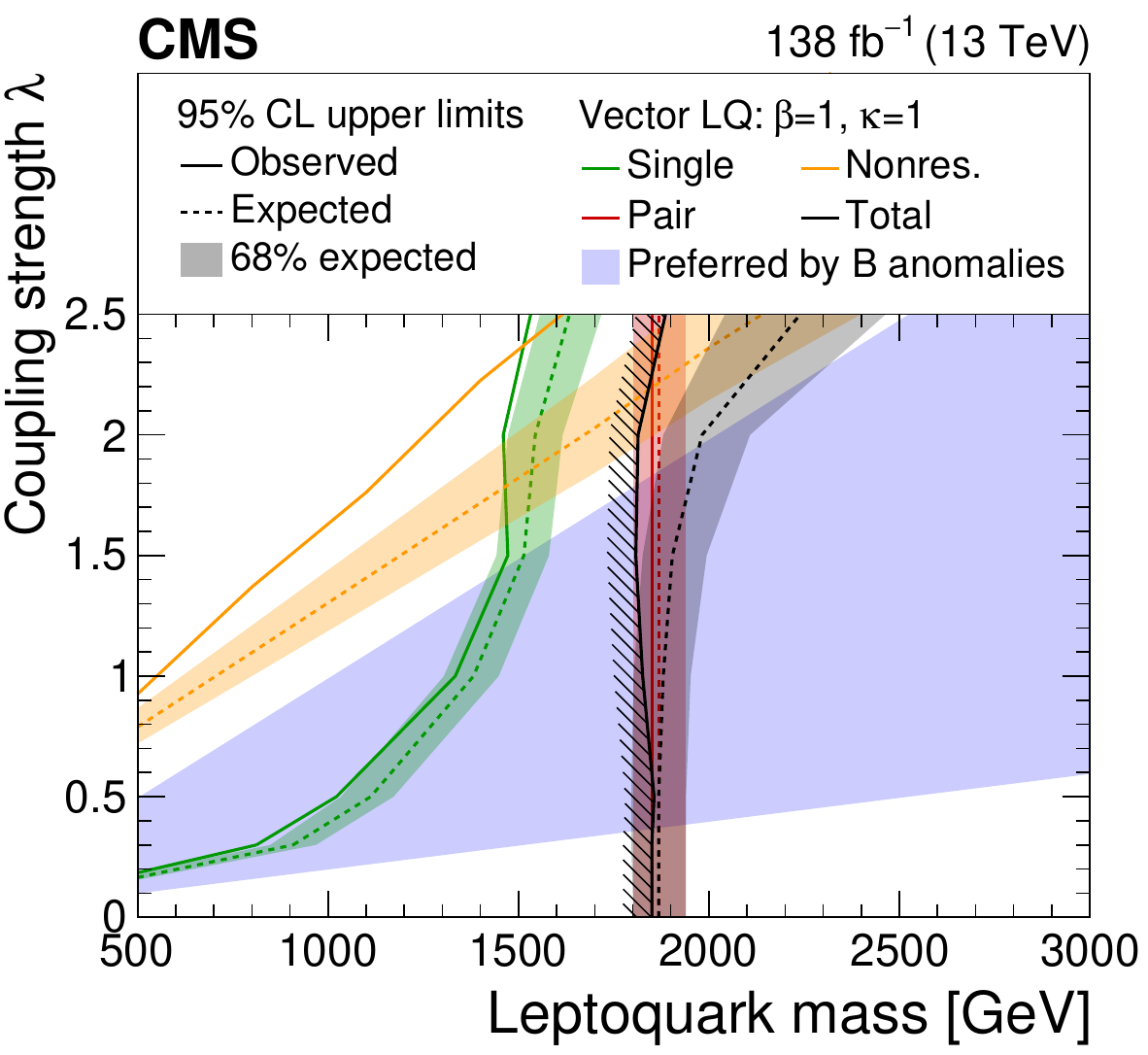}
  \caption{
Observed and expected upper limit at the 95\% \CL on the coupling strength $\lambda$ of a vector \LQ model with $\kappa=0$ (left) and $\kappa=1$ (right) under the assumption of exclusive \LQ couplings to \PQb quarks and \PGt leptons.
The limits derived for the single (green), pair (red), nonresonant (orange), and total \LQ production (black) are shown.
The shaded bands around the expected limit lines correspond to the regions containing 68\% of the distribution of limits expected under the background-only hypothesis. The hatches indicate the excluded side of the parameter space with respect to the combined observed limit.
The region with blue shading shows the parameter space preferred by one of the models proposed to explain the \PB physics anomalies~\cite{RD_LQ_pheno30}.
  }
  \label{fig:limit_combined_VLQ_2D}
\end{figure}

\begin{figure}[htb!]
  \centering
  \includegraphics[width=1.1\cmsFigWidth]{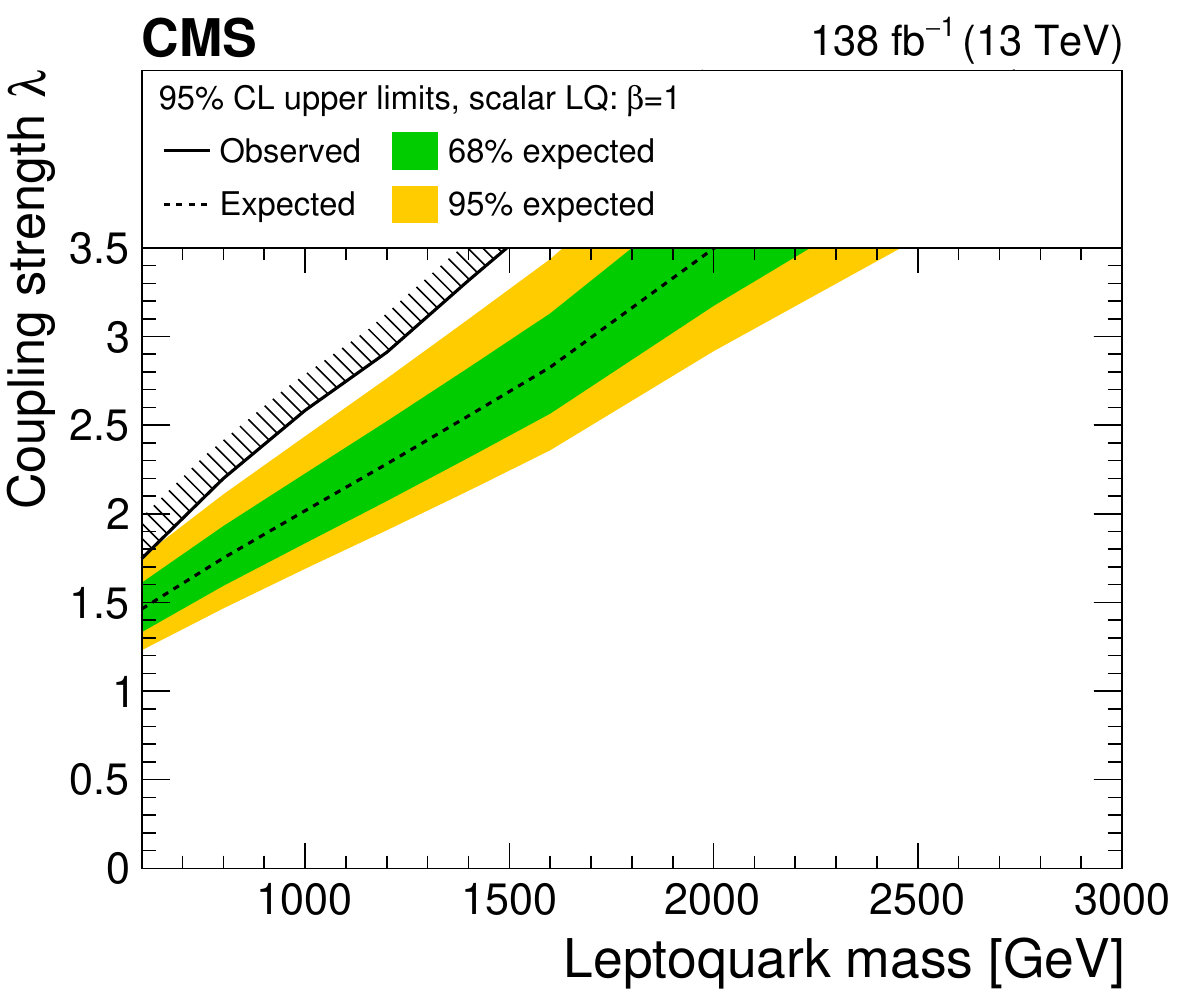}\quad
  \includegraphics[width=1.1\cmsFigWidth]{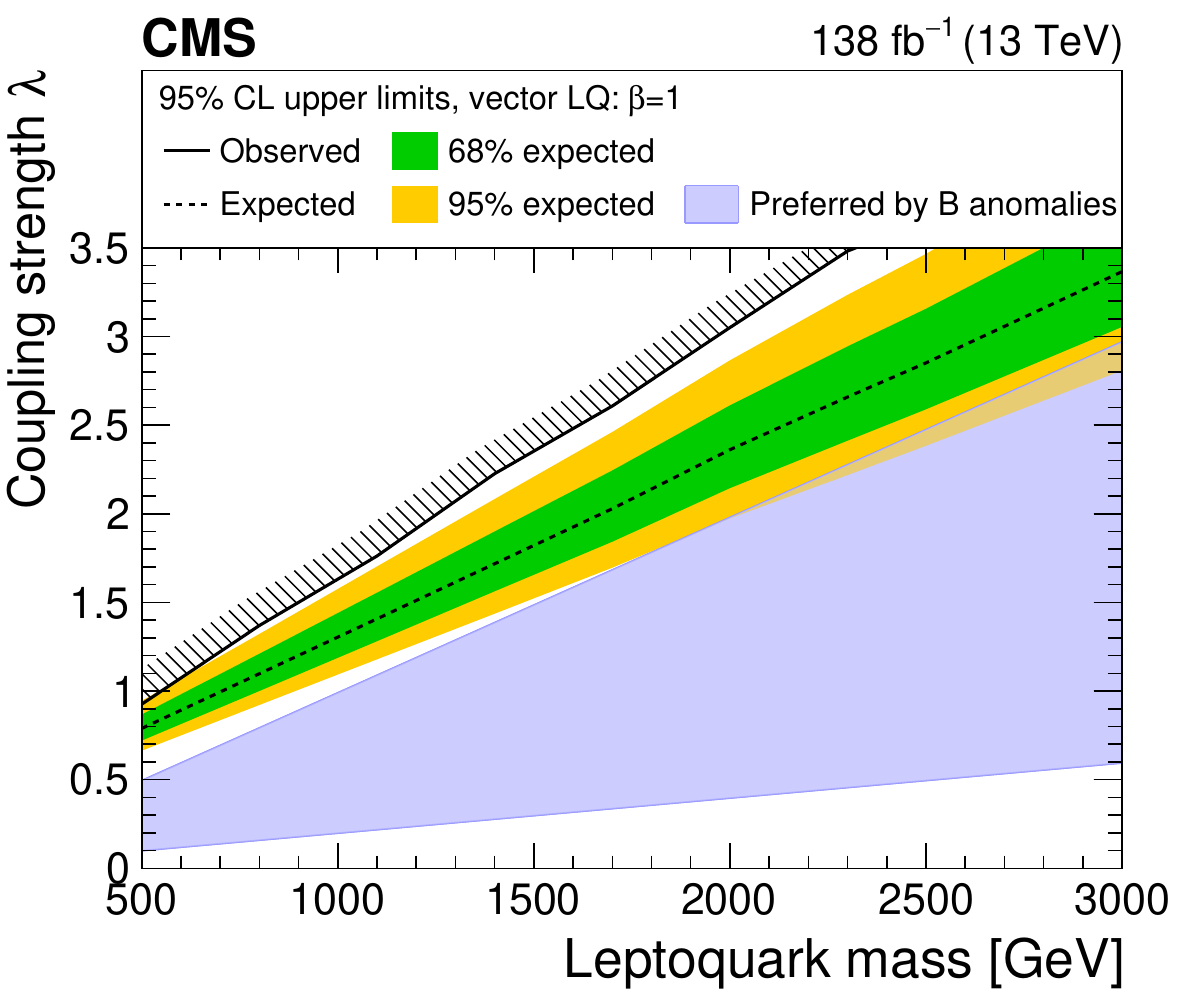}
  \caption{
Observed and expected upper limit at the 95\% \CL on the coupling strength $\lambda$ of a scalar (left) and vector \LQ model (right)
determined by considering only the nonresonant production of two \PGt leptons through $t$-channel \LQ exchange. Exclusive \LQ couplings to \PQb quarks and \PGt leptons are assumed.
The inner (green) band and the outer (yellow) band indicate the regions containing 68 and 95\%,
respectively, of the distribution of limits expected under the background-only hypothesis. The hatches indicate the excluded side of the parameter space with respect to the observed limit. The region with blue shading shows the parameter space preferred by one of the models proposed to explain the \PB physics anomalies~\cite{RD_LQ_pheno30}.
  }
  \label{fig:limit_combined_2D_nonres}
\end{figure}

Figures~\ref{fig:limit_combined_SLQ_2D} and \ref{fig:limit_combined_VLQ_2D} show the upper limit on $\lambda$ at the 95\% \CL as a function of \mLQ for the scalar and vector model. They include one line for each production mode, as well as their combination, which allows for the exclusion of a larger region at higher values of $\lambda$, where the single and nonresonant production start to contribute more significantly.
Figure~\ref{fig:limit_combined_2D_nonres} shows the upper limit on the coupling strength for \LQs exchanged in the $t$ channel with an extended coupling range, placing constraints on \LQs with masses and coupling strengths of up to 3\TeV and 3.5, respectively.
The region with blue shading shows the parameter space preferred by one of the models proposed to explain the \PB physics anomalies~\cite{RD_LQ_pheno30}.
At 95\% confidence level, \LQ masses below 1.22\TeV are excluded for the scalar model with $\lambda=1$, and below 1.50 (1.82)\TeV for the vector model with $\lambda=1$ and a coupling parameter $\kappa=0$\,(1). At $\lambda=2.5$, the lower limits are 1.31\TeV for a scalar model, and 1.73 (1.88)\TeV for a vector model with $\kappa=0$\,(1).

\clearpage

\section{Summary} \label{sec:summary}

A search has been presented for a third-generation leptoquark (\LQ) decaying to a \PGt lepton and a \PQb quark. Events with \PGt leptons and a varying number of jets originating from \PQb quarks are considered, targeting the single and pair production of the LQs, as well as the nonresonant production of two \PGt leptons through $t$-channel \LQ exchange. The search uses proton-proton collision data at a center-of-mass energy of 13\TeV recorded with the CMS detector and corresponding to an integrated luminosity of 138\fbinv. Upper limits are set on third-generation scalar and vector \LQ production cross sections as a function of \LQ mass, and results are compared with theoretical predictions to obtain lower limits on the \LQ mass. At 95\% confidence level, third-generation \LQs decaying to a \PGt lepton and a \PQb quark with unit coupling ($\lambda=1$) are excluded for masses below 1.22\TeV for a scalar model, and below 1.50 (1.82)\TeV for a vector model with a coupling parameter $\kappa=0$\,(1). For $\lambda=2.5$ the lower limits are 1.31\TeV for a scalar model, and 1.73 (1.88)\TeV for a vector model with $\kappa=0$\,(1).
The study of nonresonant \tautau production through $t$-channel \LQ exchange allows lower limits on the \LQ mass of up to 2.3\TeV to be obtained.
Upper limits are also set on the coupling strengths of scalar and vector \LQs as functions of their mass.

The observed data are found to agree with the standard model expectation within 2 standard deviations below a coupling strength of $\lambda=1.5$.
For a benchmark \LQ model with a mass of 2\TeV and a coupling strength of 2.5, the data show an excess with a local significance of 2.8 standard deviations above the standard model expectation.
Consequently, the observed upper limits on the \LQ production cross section are about three times larger than expected for this benchmark.
The present excess is driven by events with at least one highly energetic jet but no \PQb-tagged jets, indicating the need for future similar analyses to consider alternative signal models.

\begin{acknowledgments}
We congratulate our colleagues in the CERN accelerator departments for the excellent performance of the LHC and thank the technical and administrative staffs at CERN and at other CMS institutes for their contributions to the success of the CMS effort. In addition, we gratefully acknowledge the computing centers and personnel of the Worldwide LHC Computing Grid and other centers for delivering so effectively the computing infrastructure essential to our analyses. Finally, we acknowledge the enduring support for the construction and operation of the LHC, the CMS detector, and the supporting computing infrastructure provided by the following funding agencies: SC (Armenia), BMBWF and FWF (Austria); FNRS and FWO (Belgium); CNPq, CAPES, FAPERJ, FAPERGS, and FAPESP (Brazil); MES and BNSF (Bulgaria); CERN; CAS, MoST, and NSFC (China); MINCIENCIAS (Colombia); MSES and CSF (Croatia); RIF (Cyprus); SENESCYT (Ecuador); MoER, ERC PUT and ERDF (Estonia); Academy of Finland, MEC, and HIP (Finland); CEA and CNRS/IN2P3 (France); BMBF, DFG, and HGF (Germany); GSRI (Greece); NKFIH (Hungary); DAE and DST (India); IPM (Iran); SFI (Ireland); INFN (Italy); MSIP and NRF (Republic of Korea); MES (Latvia); LAS (Lithuania); MOE and UM (Malaysia); BUAP, CINVESTAV, CONACYT, LNS, SEP, and UASLP-FAI (Mexico); MOS (Montenegro); MBIE (New Zealand); PAEC (Pakistan); MES and NSC (Poland); FCT (Portugal); MESTD (Serbia); MCIN/AEI and PCTI (Spain); MOSTR (Sri Lanka); Swiss Funding Agencies (Switzerland); MST (Taipei); MHESI and NSTDA (Thailand); TUBITAK and TENMAK (Turkey); NASU (Ukraine); STFC (United Kingdom); DOE and NSF (USA).

\hyphenation{Rachada-pisek} Individuals have received support from the Marie-Curie program and the European Research Council and Horizon 2020 Grant, contract Nos.\ 675440, 724704, 752730, 758316, 765710, 824093, and COST Action CA16108 (European Union); the Leventis Foundation; the Alfred P.\ Sloan Foundation; the Alexander von Humboldt Foundation; the Science Committee, project no. 22rl-037 (Armenia); the Belgian Federal Science Policy Office; the Fonds pour la Formation \`a la Recherche dans l'Industrie et dans l'Agriculture (FRIA-Belgium); the Agentschap voor Innovatie door Wetenschap en Technologie (IWT-Belgium); the F.R.S.-FNRS and FWO (Belgium) under the ``Excellence of Science -- EOS" -- be.h project n.\ 30820817; the Beijing Municipal Science \& Technology Commission, No. Z191100007219010 and Fundamental Research Funds for the Central Universities (China); the Ministry of Education, Youth and Sports (MEYS) of the Czech Republic; the Shota Rustaveli National Science Foundation, grant FR-22-985 (Georgia); the Deutsche Forschungsgemeinschaft (DFG), under Germany's Excellence Strategy -- EXC 2121 ``Quantum Universe" -- 390833306, and under project number 400140256 - GRK2497; the Hellenic Foundation for Research and Innovation (HFRI), Project Number 2288 (Greece); the Hungarian Academy of Sciences, the New National Excellence Program - \'UNKP, the NKFIH research grants K 124845, K 124850, K 128713, K 128786, K 129058, K 131991, K 133046, K 138136, K 143460, K 143477, 2020-2.2.1-ED-2021-00181, and TKP2021-NKTA-64 (Hungary); the Council of Science and Industrial Research, India; the Latvian Council of Science; the Ministry of Education and Science, project no. 2022/WK/14, and the National Science Center, contracts Opus 2021/41/B/ST2/01369 and 2021/43/B/ST2/01552 (Poland); the Funda\c{c}\~ao para a Ci\^encia e a Tecnologia, grant CEECIND/01334/2018 (Portugal); the National Priorities Research Program by Qatar National Research Fund;  MCIN/AEI/10.13039/501100011033, ERDF ``a way of making Europe", and the Programa Estatal de Fomento de la Investigaci{\'o}n Cient{\'i}fica y T{\'e}cnica de Excelencia Mar\'{\i}a de Maeztu, grant MDM-2017-0765 and Programa Severo Ochoa del Principado de Asturias (Spain); the Chulalongkorn Academic into Its 2nd Century Project Advancement Project, and the National Science, Research and Innovation Fund via the Program Management Unit for Human Resources \& Institutional Development, Research and Innovation, grant B05F650021 (Thailand); the Kavli Foundation; the Nvidia Corporation; the SuperMicro Corporation; the Welch Foundation, contract C-1845; and the Weston Havens Foundation (USA).
\end{acknowledgments}

\bibliography{auto_generated}
\cleardoublepage \appendix\section{The CMS Collaboration \label{app:collab}}\begin{sloppypar}\hyphenpenalty=5000\widowpenalty=500\clubpenalty=5000
\cmsinstitute{Yerevan Physics Institute, Yerevan, Armenia}
{\tolerance=6000
A.~Hayrapetyan, A.~Tumasyan\cmsAuthorMark{1}\cmsorcid{0009-0000-0684-6742}
\par}
\cmsinstitute{Institut f\"{u}r Hochenergiephysik, Vienna, Austria}
{\tolerance=6000
W.~Adam\cmsorcid{0000-0001-9099-4341}, J.W.~Andrejkovic, T.~Bergauer\cmsorcid{0000-0002-5786-0293}, S.~Chatterjee\cmsorcid{0000-0003-2660-0349}, K.~Damanakis\cmsorcid{0000-0001-5389-2872}, M.~Dragicevic\cmsorcid{0000-0003-1967-6783}, A.~Escalante~Del~Valle\cmsorcid{0000-0002-9702-6359}, P.S.~Hussain\cmsorcid{0000-0002-4825-5278}, M.~Jeitler\cmsAuthorMark{2}\cmsorcid{0000-0002-5141-9560}, N.~Krammer\cmsorcid{0000-0002-0548-0985}, D.~Liko\cmsorcid{0000-0002-3380-473X}, I.~Mikulec\cmsorcid{0000-0003-0385-2746}, J.~Schieck\cmsAuthorMark{2}\cmsorcid{0000-0002-1058-8093}, R.~Sch\"{o}fbeck\cmsorcid{0000-0002-2332-8784}, D.~Schwarz\cmsorcid{0000-0002-3821-7331}, M.~Sonawane\cmsorcid{0000-0003-0510-7010}, S.~Templ\cmsorcid{0000-0003-3137-5692}, W.~Waltenberger\cmsorcid{0000-0002-6215-7228}, C.-E.~Wulz\cmsAuthorMark{2}\cmsorcid{0000-0001-9226-5812}
% ...
\par}
\vskip\cmsinstskip
\dag:~Deceased\\
$^{1}$Also at Yerevan State University, Yerevan, Armenia\\
$^{2}$Also at TU Wien, Vienna, Austria\\
% ...
\end{sloppypar}
\end{document}